\documentclass[review]{elsarticle}

\usepackage{lineno,hyperref}

\hypersetup{
    bookmarks=true,         
    unicode=false,          
    pdftoolbar=true,        
    pdfmenubar=true,        
    pdffitwindow=false,     
    pdftitle={Subgraph Isomorphism in Temporal Networks},    
    pdfauthor={Ursula Redmond},     
    pdfsubject={Subgraph Isomorphism in Temporal Networks},   
    pdfcreator={Ursula Redmond},   
    pdfproducer={},  
    pdfkeywords={network analysis} {temporal networks} {graph mining}, 
    pdfnewwindow=true,      
    colorlinks=false,       
    linkcolor=red,          
    citecolor=green,        
    filecolor=magenta,      
    urlcolor=cyan           
}

\makeatletter
\def\ps@pprintTitle{%
  \let\@oddhead\@empty
  \let\@evenhead\@empty
  \let\@oddfoot\@empty
  \let\@evenfoot\@oddfoot
}
\makeatother

\modulolinenumbers[5]

\usepackage{algorithm}
\usepackage{algorithmic}
\usepackage{subfig}

\usepackage{epstopdf}
\usepackage{multirow}
\usepackage{numprint}
\usepackage{graphicx}
\usepackage{verbatim}
\newcommand{\vf}{To-Ti}
\newcommand{\tempvf}{Ti\&To}
\newtheorem{definition}{Definition}
\biboptions{authoryear}

\begin{document}

\begin{frontmatter}

\title{Subgraph Isomorphism in Temporal Networks}



\author[mymainaddress]{Ursula Redmond\corref{mycorrespondingauthor}}
\cortext[mycorrespondingauthor]{Corresponding author}
\ead{u.redmond@gmail.com}

\author[mymainaddress]{P\'{a}draig Cunningham}
\ead{padraig.cunningham@ucd.ie}

\address[mymainaddress]{School of Computer Science and Informatics, University College Dublin, Ireland}

\begin{abstract}
Temporal information is increasingly available as part of large network data sets. This information reveals sequences of link activations between network entities, which can expose underlying processes in the data. Examples include the dissemination of information through a social network, the propagation of musical ideas in a music sampling network, and the spread of a disease via contacts between infected and susceptible individuals. The search for these more meaningful patterns may be formulated as a time-respecting subgraph isomorphism problem.  Our set of query graphs include an enumeration of small random graphs and fan-out-fan-in structures, all composed of time-respecting paths. We explore three methods of solving the problem, which differ in how they exploit temporal and topological information. One approach extracts all subgraphs that have the temporal properties we require and then performs subgraph isomorphism testing on each subgraph. Another approach performs subgraph isomorphism testing first with temporal post-filtering, while the other is a hybrid approach that uses temporal information during the search. We empirically demonstrate the hybrid approach to be more efficient than the others, over a range of network data sets. These data come from communication and social networks, up to \numprint{20000} interactions in size.
\end{abstract}


\end{frontmatter}



\section{Introduction}

Increasingly, network data sets include information describing the timing of connections between entities. This can reveal features of the data that are impossible to unearth from the static version of a network. The data encode sequences of events that follow particular patterns, which represent different types of propagation, depending on the application. For example, in a communication network, the timing of interactions can illustrate the flow of a piece of information. In a network of micro-blogs, the time at which links occur can reflect the diffusion of sentiment. Infection spreads among individuals through their physical contacts, and depends on features such as the time of infection, the interval during which an individual is infectious, and the time at which they recover. In a network of financial transactions on peer-to-peer lending websites, a temporally increasing path of loans could indicate the presence of arbitrageurs \citep{redmond-13}.

Traditionally in temporal network analysis, the time slicing approach first uses a sliding window to isolate data that occur between a start and end time, then examines the resultant data as if it were static. This risks arbitrarily cutting potential contagions in a network, such as when the originating and terminating individuals are not present in a time-slice, as a result of the time window choice. To avoid this problem, we use temporal constraints at the interaction level. In this way, given a contagion in the network, it can be found if its pair-wise interactions obey the temporal constraints.

Given a particular query graph pattern of interest, our task is to find all of its embeddings in a given network graph, such that the embeddings adhere to temporal constraints. We approach this as a subgraph isomorphism problem, known to be NP-complete \citep{garey-johnson-79}, and  use temporal information to guide the search. While quite disparate approaches to modelling networks in the temporal domain exist, \citep{holme-12}, our model of temporally constrained query embeddings captures event sequences which can be meaningfully interpreted by domain experts.

This paper further develops our initial work \citep{redmond-asonam-13}. We present a data-driven method for selecting the time-delay threshold for interactions between individuals. We apply our subgraph matching methods to a wide range of networks, and describe ways to interpret query embeddings according to the phenomena represented by each network.

We propose a set of comparisons to demonstrate the benefits associated with the active use of temporal information for subgraph matching. The three methods are differentiated by their use of temporal and topological information at different stages of the subgraph matching algorithm, and named accordingly.

One approach, \emph{Time before Topology}, extracts all time-respecting subgraphs from the target network. The subgraph isomorphism test is then applied to each extracted subgraph in turn. The extraction step is very time-consuming, and since the resultant subgraphs can have a large degree of overlap, there may be very many subgraph isomorphism tests, some of which are redundant.

Another approach, \emph{Topology before Time}, works the other way around. First, the subgraph isomorphism test is performed on the entire network, without leveraging temporal information, thus returning all static embeddings. Then, the temporal information associated with each embedding is checked. Only embeddings which are time-respecting are retained.

The method we show to be most effective, \emph{Time and Topology Together}, lies between these two approaches. At each step of the subgraph isomorphism test being computed, temporal information associated with the interactions is used to prune the search space. This results in fewer candidate matches, and correspondingly to fewer time-related checks.

In this paper, we describe these algorithms, and empirically compare their relative performances. We demonstrate that the amount of pruning achieved with the most efficient approach yields significant performance gains. Section \ref{sec:related} presents related work. Section \ref{sec:methods} introduces our theoretical framework, formalizing the required ideas and presenting our algorithmic comparisons. The data sets we use are described in Section \ref{sec:data}. Section \ref{sec:results} presents the results of applying our algorithms to a collection of network data sets. Section \ref{sec:conc} concludes the paper with a discussion and suggestions for future work.


\section{Related Work}\label{sec:related}

This work draws on topics from a variety of network analysis fields. Related work includes the long-studied problem of subgraph isomorphism, as well as the more recently popular areas of temporal network and cascade analysis.

\subsection{Subgraph Isomorphism}

The subgraph isomorphism problem determines whether a given graph contains a subgraph which has the same topological structure as another given graph. Subgraph isomorphism is an NP-complete problem \citep{garey-johnson-79}. Thus, the time complexity of brute force matching algorithms increases exponentially with the size of the graphs and query graphs to be matched. This makes the problem prohibitively expensive to solve for graphs of large scale.

A backtracking algorithm was proposed by Ullmann to solve the graph and subgraph isomorphism problems \citep{ullmann-76}. In an extension to the algorithm, the search space is pruned based on the degree of nodes in the graphs to be matched. Another approach \citep{schmidt-76} also employs backtracking, but uses the distance matrix representation of a graph to inspire the pruning steps.

The VF algorithm of Cordella \emph{et al.} presents a depth-first search strategy for graph and subgraph isomorphism \citep{cordella-et-al-99}. The matching process is described by a state space representation, in which each state of the process is associated with a partial solution. The partial solution includes the elements of the two graphs which match each other so far. This partial solution is expanded at the next step, and the algorithm evaluates whether a full match has been found. If not, the search continues. The speed of the algorithm compares favourably with Ullmann's often-used backtracking approach. An enhanced version, VF2, provides further performance gains \citep{cordella-et-al-01} by substantially reducing memory requirements.

Both VF and VF2 are presented as recursive algorithms. The improvements with VF2 come partly from the data structures it maintains at each level of recursion, which reduce the amount of computation required. Since our problem is in a temporal network setting, we must also consider temporal information at each stage of the matching process. We adopt a recursive approach, and maintain data structures which account for topological and temporal information. Since our proposed solutions are achieved with a similar methodology to VF2, we use VF2 as a comparison to evaluate the relative merits of our approach.

Graph simulation has been extended to tackle the problem of graph pattern matching. Ma \emph{et al.} propose a version of graph simulation which preserves the topology of data graphs, and limits the number of matches \citep{ma-11}. The authors also introduce an algorithm to perform the computation, and experimentally verify its effectiveness on real-life and synthetic data. Since many modern networks are large and frequently updated with changes, Fan \emph{et al.} propose an incremental graph pattern matching algorithm \citep{fan-13} which also employs graph simulation. The authors avoids recomputing matches from scratch when the data graph is updated. A subgraph isomorphism method was proposed in the realm of biochemical data \citep{bonnici-13}. The authors present their approach, along with a comparison to other methods in order to demonstrate its value. Their algorithm creates a search strategy based on the topology of the query graph, and uses this to prune the search space as early as possible.

Another approach, \emph{AllDifferent}-based filtering, was introduced \citep{solnon-10} to tackle the problem from a constraint satisfaction point of view. This is achieved by associating a variable with every node in the query graph. Solving the problem then involves finding assignments of the variables that satisfy the matching constraints. Ullmann also proposed a constraint satisfaction approach, which utilizes bit-vectors \citep{ullmann-10}.

Subgraph isomorphism is commonly associated with finding a query graph from a large set of graphs, which may be stored in a database. This problem has been widely addressed, with some approaches focussed on filtering-and-verification. In the first step, the set of graphs is filtered, and only the graphs which are likely to contain the query are retrieved. In the second step, the more costly verification is performed on these likely candidates. A filtering-and-verification approach was presented \citep{shang-08}. The authors propose an efficient algorithm for testing subgraph isomorphism, and a feature-based technique for indexing the database to allow for quick filtering and to work well with their subgraph isomorphism algorithm. The GString algorithm was presented \citep{jiang-07}, which represents graphs and queries by sequences, converting graph search to subsequence matching. The authors use meaningful graph structure components as the most basic units of sequencing, leading to a more efficient implementation. The approach is presented in terms of chemical compound databases, so the basic units of sequencing are chemical compounds understood by domain experts.

He \emph{et al.} introduce the notion of a graph closure -- a generalised graph to represent a set of graphs -- and use it to construct an index for graph queries \citep{he-06}. Their \emph{Closure-tree} indexing technique organizes graphs hierarchically, with each node summarizing its descendants using a graph closure. In another work, web-scale graph data is tackled \citep{sun-12}. The authors present an algorithm to support subgraph matching for distributed graphs. The algorithm uses graph exploration and parallel computing to process queries. The problem of subgraph pattern search over graph streams has been explored \citep{chen-10}. The problem takes the form of a continuous join between the query pattern and the graph stream, the join predicate being a subgraph isomorphism. Since the subgraph isomorphism problem is NP-complete, the authors propose an approximate solution to the problem.

Other approaches have put constraints on the topology of the query graph or the network graph. Matula focussed on subtree isomorphism \citep{matula-78}, showing that the act of checking if a query tree is isomorphic any subtree of a larger tree can be achieved in a bounded amount of steps. Trees and forests are outerplanar structures. Sys\l{}o showed that the subgraph isomorphism problem for outerplanar graphs  remains NP-complete, even with strong connectivity constraints placed on both the query and data graph \citep{syslo-82}. Horv\'{a}th \emph{et al.} introduce frequent subgraph mining in outerplanar graphs \citep{horvath-10}. The authors demonstrate their approach to work in incremental polynomial time for outerplanar graphs which have only polynomially many simple cycles. They perform the analysis on network data from chemoinformatics and bioinformatics applications.
 
\subsection{Temporal Network Analysis}

Many concepts developed for static graphs must be revised to account for temporal information. A comprehensive review \citep{holme-12} details these concepts. A \emph{time-respecting} path is defined as a sequence of contacts with non-decreasing times \citep{kempe-02}. Nodes $i$ and $j$ are \emph{strongly connected} if there is a time-respecting, directed path connecting $i$ to $j$ and vice versa \citep{nicosia-et-al-12}. Nodes $i$ and $j$ are \emph{weakly connected} if there are time-respecting undirected paths from $i$ to $j$ and vice versa.

For a directed edge to exist between nodes $i$ and $j$ in a \emph{reachability graph}, there must be a time-respecting path between them. Reachability graphs reveal the nodes which are reachable from a single root node \citep{moody-02}. Analysis of the reachability graph within a dating network of high-school students reveals interesting relationships \citep{bearman-04}. Reachability graphs over time are studied in networks from Facebook, Twitter, and the Enron email data set \citep{macropol-12}. These structures show all nodes that can be influenced by a specific root node over different time scales. A \emph{time-respecting subgraph} \citep{redmond-13} is a generalization of a reachability graph, since it does not require a root node, but insists on reachability along each directed path.

The lifespan of a piece of information in a temporal communication network may be specified by a time window \citep{zhao-10} This measures the time between the end of one communication and the beginning of the next. The idea is that the closer in time the contacts take place, the more likely they are to discuss the same topic. Similarly, the \emph{relay time} of an interaction captures the time taken for a newly infected node to spread the infection further via the next interaction that the link participates in \citep{kivela-12}. We incorporate this feature when examining time-respecting subgraphs, requiring that consecutive interactions occur within a specified time. The spread of information through a network can also be modeled by a cascade. The structure of cascades can reveal trends in a network that facilitate spreading and community development \citep{ghosh-11}. The importance of \emph{time-constrained} cascades is emphasized for understanding contagion \citep{banos-13}.

Temporal motifs are connected subgraphs composed of similar event sequences, where similarity is measured in terms of the topology and temporal ordering of the events \citep{kovanen-11}. All adjacent events in a temporal motif must occur within time $\Delta t$ of each other. The  events connected to a node must be consecutive in time. So if a node $n$ in a temporal motif participates in events at times $t_{0}$ and $t_{2}$, then if an event exists involving $n$ at time $t_{1}$, it must be included in the motif so that the motif is valid. This is distinct from a flow motif, in which directed events that meet head-to-tail must be consecutive in time. Kovanen \emph{et al.} propose an algorithm to find temporal motifs, which do not have the flow requirement. The aim of our approach is to efficiently find a time-respecting subgraph that is of interest to an expert user in a temporal network. This means that interactions in the subgraph occur within a specified time of each other, and that events meeting head-to-tail must be consecutive in time.

As motif analysis does for static networks, temporal motif analysis reveals much about the underlying mechanisms within temporal networks. In subsequent work \citep{kovanen-13}, temporal motifs in a mobile communication network are explored. By including other attributes of the data, further interesting mechanisms were found within the temporal motifs. These included gender-related differences in communication patterns, and a tendency for similar individuals to communicate more often than might be expected.


\section{Methods}\label{sec:methods}

This section formally describes our problem framework. We outline our three proposed solutions to the temporal subgraph matching problem. To determine the most effective approach, we compare the three algorithms, each one making use of temporal and topological information in separate phases.

\subsection{The Problem Framework}

In all of our algorithms, we must solve the subgraph isomorphism problem to find specified query graphs in a temporal network. The definition of subgraph isomorphism may be presented as follows \citep{ullmann-76}: 


\begin{definition}
A graph $G2$ is isomorphic to a subgraph of a graph $G1$ if and only if there is a one-to-one correspondence between the node sets of this subgraph and of $G2$ that preserves adjacency.
\end{definition}

Since time is encoded explicitly as part of our network representation, instead of referring to an ``edge" between two nodes, we use the term ``interaction" to specify a triplet, made up of two nodes and the time of their contact. We define a directed temporal graph as follows:


\begin{definition}
A directed temporal graph $G$ consists of a set $V$ of nodes and a set $E$ of ordered pairs of nodes representing interactions. An interaction $e_{i} \in E$ is represented by a three-tuple $e_{i} = (u_{i}, v_{i}, t_{i})$, in which $u_{i}$ is the source node, $v_{i}$ is the target node and $t_{i}$ is the initiation time of the interaction.
\end{definition}


In a directed temporal network, we have the notion of a contagion being initiated by an individual and spreading over a sequence of interactions to others in the network. In order for a contagion to take place, adjacent interactions must be time-respecting.

\begin{definition}
Let $e_{i} = (u_{i}, v_{i}, t_{i})$ and $e_{j} = (u_{j}, v_{j}, t_{j})$ be interactions in a directed temporal graph. Given some threshold $d$, the interactions are time-respecting if they are adjacent and:

$\left\{
	\begin{array}{ll}
		0 \leq t_{j} - t_{i} \leq d  & \mbox{if } v_{i} = u_{j} \\
		0 \leq t_{i} - t_{j} \leq d & \mbox{if } u_{i} = v_{j} \\
		0 \leq | t_{j} - t_{i} | \leq d & \mbox{otherwise }
	\end{array}
\right.$
\label{def:tr}
\end{definition}

In the first two cases of Definition \ref{def:tr}, The interactions meet head-to-tail. In the final case, the interactions either share a source node or a target node. This relation between time-respecting interactions is a constraint, which we incorporate into our definition of a time-respecting path and subgraph.

Time-respecting paths describe a non-decreasing sequence of interactions \citep{pan-11}. A path can be thought of as a mechanism for passing information from a source, along a sequence of intermediaries, to a target. This serves to emphasize our view of the network from a temporal standpoint. With traditional time-slicing, the specified time window determines which interactions are examined, between a minimum and maximum interaction time. However, a time-respecting path has no such bounds. In fact, given the right connectivity and timing of interactions, a path might be initiated when the network is first created, and continue until the latest point in the data set.


We define a time-respecting subgraph in terms of time-respecting interactions. In this work, since the query graphs are connected, we require that the embedded subgraphs are connected.

\begin{definition}
A time-respecting subgraph $S = (V', E')$ of a temporal graph $G = (V, E)$ is a subgraph such that every adjacent interaction pair in $E'$ is time-respecting.
\label{def:trsgs}
\end{definition}

It is important to note that in our implementation, embedded subgraphs are induced. Thus, given any pair of nodes in an embedded subgraph, all interactions between them are included in the embedding. So, if a potential embedding includes more interactions than specified by the query graph, the embedding will not be returned.


\subsection{Time Before Topology}

\begin{figure}[]
	\centering
	\includegraphics[width=0.3\textwidth, height = 0.11\textheight]{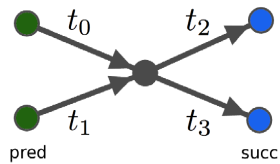}
\caption{An example of a time-respecting subgraph. Here, $t_{0} \leq t_{1} \leq t_{2} \leq t_{3}$. All interactions which are incident to the same node must occur within time $d$ of each other. Thus, we require that $| t_{3} - t_{0} | \leq d$. All incoming edges to a node $n$ must precede all outgoing edges from node $n$. So, we must have that $t_{0} \leq t_{2}$, $t_{0} \leq t_{3}$, $t_{1} \leq t_{2}$ and $t_{1} \leq t_{3}$.}
\label{fig:pairs}
\end{figure}

This approach addresses the temporal dimension of the problem first. Since the matched embeddings returned must be time-respecting subgraphs, we first extract all possible time-respecting subgraphs from the network. The algorithm we use to do this is similar to the well known method for finding maximal connected components in graphs \citep{hopcroft-73}. In our setting, we use a breadth-first search on the interactions, rather than on the nodes. This is motivated by the fact that each node may be involved in multiple sets of interactions at different times. The search begins at an interaction $e$ and expands until the maximal time-respecting subgraph containing $e$ is found. Each adjacent interaction pair in a subgraph must fulfil the requirements illustrated in \figurename{\ref{fig:pairs}} in order to be time-respecting. A new search is performed from each interaction not already included in a time-respecting subgraph.

This first step returns the set of all possible time-respecting subgraphs in the given network. The subgraph isomorphism test is then applied to each extracted time-respecting subgraph in turn (ignoring time-stamps) using the VF2 algorithm, and positive matches are retained. The time-stamps may be safely ignored, since any subgraph of a time-respecting subgraph will itself be time-respecting.


\begin{figure}[]
\centering
 
\subfloat[The network G1 and a query graph G2.]{
	\includegraphics[width=0.7\textwidth, height = 0.25\textheight]{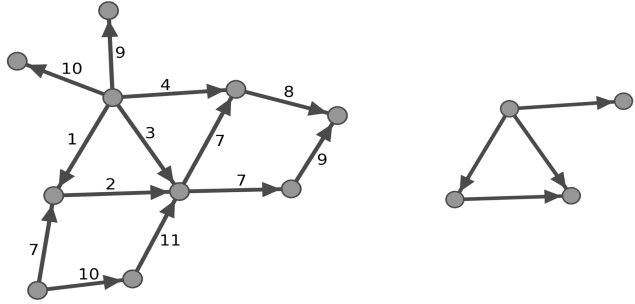}
	\label{fig:G1G2}}\\
\subfloat[All maximal time-respecting subgraphs, extracted from G1. At each node, interactions occur within time $d$ = 4 of each other, and incoming interactions precede outgoing interactions.]{
	\includegraphics[width=0.65\textwidth, height = 0.277\textheight]{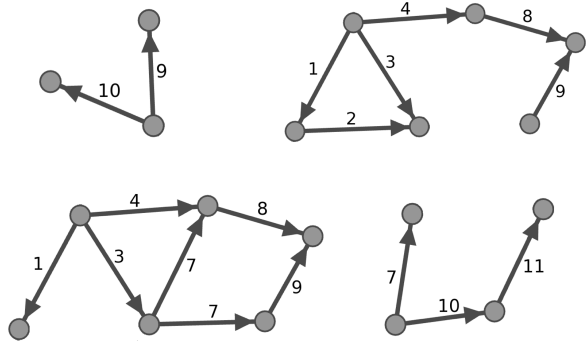}
	\label{fig:time}}\\
\subfloat[The two time-respecting embeddings of G2 in G1, found by searching for an embedding of G2 in each of the maximal time-respecting subgraphs found in the previous step.]{
	\includegraphics[width=0.52\textwidth, height = 0.11\textheight]{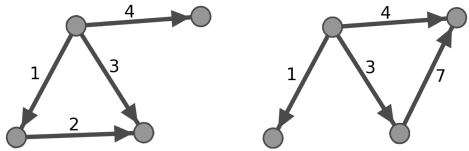}
	\label{fig:solutions}}

\caption{The \emph{Time before Topology} method. (\ref{fig:G1G2}) Given a network G1 and a query graph G2, we aim to find all time-respecting embeddings of G2 in G1. (\ref{fig:time}) First, we extract all maximal time-respecting subgraphs from G1. We then ignore the time-stamps on the interactions of the time-respecting subgraphs. We search for an embedding of G2 in each one of the time-respecting subgraphs. Note that the two subgraphs containing triangles have many overlapping components, which occurs often in larger examples and slows down the algorithm. (\ref{fig:solutions}) Two embeddings are found.}
\label{fig:schematic_time}
\end{figure}

Ideally, the initial extraction of time-respecting subgraphs should prune the search space for the subsequent tests. This would happen if large portions of the network were not time-respecting, since those portions would not be searched. In reality, the extraction of time-respecting subgraphs can extend the search space. This happens when a part of the graph can belong to multiple time-respecting subgraphs. Thus, the set of time-respecting subgraphs may contain a high degree of overlap. If the part of the graph contained in multiple time-respecting subgraphs has the topological structure sought in the subgraph isomorphism step, it will be returned multiple times (once for each time-respecting subgraph it is embedded in). This process is illustrated in \figurename{\ref{fig:schematic_time}}.

For visual simplicity, the example network in \figurename{\ref{fig:schematic_time}} contains single interactions between nodes. In real networks, multiple interactions may occur between a pair of nodes at different times. However, since the query graphs we specify have at most one interaction between nodes, and the embedded subgraphs returned by the matching process are induced, this simple example is a good proxy for the way in which results are returned by our algorithms.


\subsection{Topology Before Time ({\vf})}

This solution first generates candidate subgraphs that match the topology of the query graph, using the VF2 subgraph isomorphism algorithm. The candidate embeddings are induced, so if one interaction occurs between a pair of nodes in the query graph, then there can only be one interaction between the corresponding nodes in the embedding.

\begin{figure}[]
\centering
 
\subfloat[The network G1 and a query graph G2.]{
	\includegraphics[width=0.7\textwidth, height = 0.25\textheight]{figures/G1G2.eps}
	\label{fig:G1G2_top}}\\
\subfloat[All topological matches of G2 found in G1.]{
	\includegraphics[width=0.8\textwidth, height = 0.4\textheight]{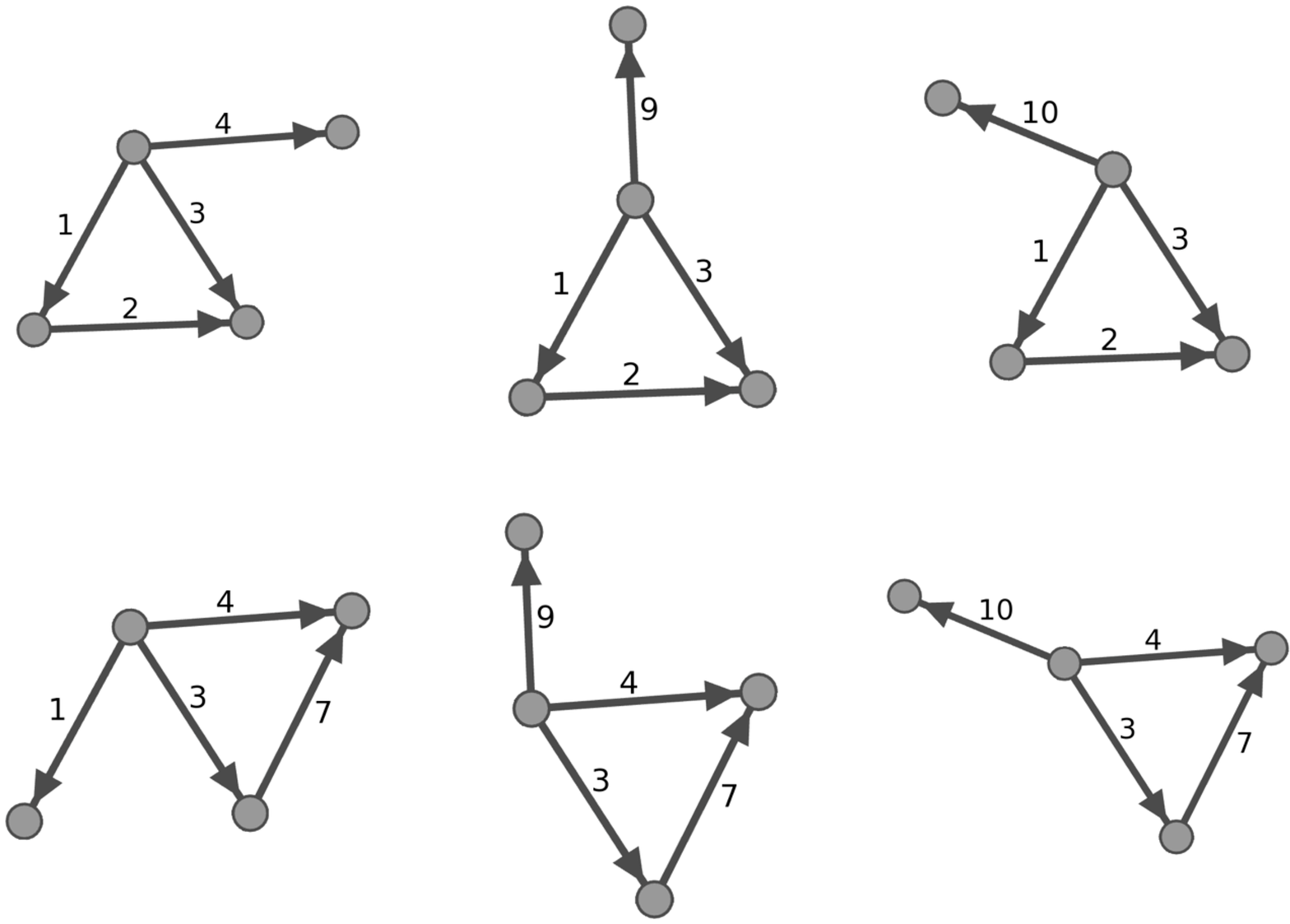}
	\label{fig:top}}\\
\subfloat[Only the matches which are time-respecting are retained.]{
	\includegraphics[width=0.5\textwidth, height = 0.11\textheight]{figures/Solutions.eps}
	\label{fig:solutions_top}}

\caption{The {\vf} method. (\ref{fig:G1G2_top}) Given a network G1 and a query graph G2, we aim to find all embeddings of G2 in G1 which are time-respecting. (\ref{fig:top}) First, we ignore the time-stamps on the interactions of G1 and extract all subgraphs which are isomorphic to G2. (\ref{fig:solutions_top}) We retain the embedded subgraphs which are also time-respecting. Note that the solutions are the same as those found in \figurename{\ref{fig:schematic_time}}.}
\label{fig:schematic_top}
\end{figure}

The returned candidates are then evaluated, to test whether the time-respecting property holds. Only references to those which are time-respecting are retained. Algorithm \ref{alg1} checks that the time difference between the earliest ($dates$.first)and latest ($dates$.last) interaction incident to a node $n$ is bounded by the threshold $d$. Algorithm \ref{alg2} checks that all incoming interactions ($pred\_dates$) to a node $n$ precede all outgoing interactions ($succ\_dates$) from $n$. A visual explanation of the tests is presented in \figurename{\ref{fig:pairs}}. An example of the process is illustrated in \figurename{\ref{fig:schematic_top}}.

This method was chosen since it enables a worthwhile comparison with our best approach, which we describe in the next subsection. The comparison highlights the performance gains in using temporal information for pruning at an earlier stage in the process. This solution also succeeds in producing results within a reasonable time, thereby facilitating an empirical comparison.

\begin{algorithm}
\caption{Test-one ($G2\_embedding, d$)}
\label{alg1}
	\begin{algorithmic}
		\FOR{$n \in G2\_embedding$.nodes()}
			\IF{$dates$.last$- dates$.first $> d$}
				\RETURN $False$
			\ENDIF
		\ENDFOR
	\RETURN $True$
	\end{algorithmic}
\end{algorithm}

\begin{algorithm}
\caption{Test-two ($G2\_embedding, d$)}
\label{alg2}
	\begin{algorithmic}
		\FOR{$n \in G2\_embedding$.nodes()}
			\IF{$succ\_dates$.first $ <pred\_dates$.last}
				\RETURN $False$
			\ENDIF
		\ENDFOR
	\RETURN $True$
	\end{algorithmic}
\end{algorithm}


\newpage
\subsection{Time and Topology Together (\tempvf)}

For the sake of consistency, we retain the notation used in the description of the recursive VF2 algorithm by Cordella \emph{et al.} \citep{cordella-et-al-01}. The implementation expanded upon in this paper is available \citep{vf2}. The matching process is described by a state space representation, in which each state $s$ of the process describes a partial mapping solution. In a state $s$, a portion of the query graph $G2$ matches a portion of the network graph $G1$. It is important to note that the portion of $G1$ in the mapping is induced. So, given a set of nodes in the mapping, any interactions between them are also present in the mapping.

Given such an intermediate state $s$, the mapping is extended by first computing candidate node pairs (a node from $G1$ and a node from $G2$). The candidate node from $G2$ is selected from the neighbours of the nodes in $G2$ that are currently in the mapping, so it is connected to the portion of the query graph currently matched. The candidate node from $G1$ is selected in the same way, from the neighbours of the nodes currently matched in the embedding from $G1$. Once the new nodes are included in the mapping, all interactions between them are also included. The two new, extended portions in the mapping must be graph isomorphic to be considered a feasible match. If they are not graph isomorphic, the candidate nodes are discarded as a matching pair, and the process continues with a new node pair.

If a topological match is confirmed, there is an option to test for semantic feasibility. This requires that attributes of the nodes or interactions being paired are compatible, as specified by the application. In our setting, we utilize the dates on which the interactions occur. Given an embedding of the subgraph $G2$ in the graph $G1$, we don't require that the dates on each paired interaction match each other, but rather that inclusion in their respective subgraph will maintain the time-respecting property of that subgraph. Since we are not interested in the actual times at which interactions occurred in the network data, only the semantic feasibility of $G1$ is checked.

The memory requirements of the VF2 algorithm are constrained partly through the use of data structures which are maintained at each level of recursion. We exploit this observation in our implementation by keeping track of both topological and temporal information in the same way. A map data structure named $core\_1$ contains the nodes in the current mapping from $G1$ to $G2$. This provides an efficient way for us to test that a candidate node for inclusion in the mapping will maintain the time-respecting property we require for all of the induced edges.

Before testing the legitimacy of a candidate node $G1\_node$, we construct a set of data structures. The list $pred$ contains the predecessors of $G1\_node$ in $G1$ which are also in $core\_1$, and thus part of the current mapping. Analogously, $succ$ contains the successors of $G1\_node$ in $G1$ which are also in $core\_1$. The lists $pred\_dates$ and $succ\_dates$ contain the dates, in increasing order, on which connections between $G1\_node$ and the relevant predecessor or successor nodes, respectively, were made. The list $dates$ combines these dates, sorted in increasing order.

As described in Definition \ref{def:trsgs}, a pairwise comparison of adjacent interactions must ensure that each pair is time-respecting. Accordingly, a candidate node must fulfil these criteria when included in a potential embedding of $G2$ in $G1$. The tests are outlined in Algorithms \ref{alg1} and \ref{alg2}, and illustrated in \figurename{\ref{fig:pairs}}.


\section{Network Data}\label{sec:data}

Most of the networks used in this evaluation come from KONECT, the Koblenz Network Collection \citep{kunegis2013handbook}. The exceptions are the Hagelloch measels data set \citep{hagelloch}, and the WhoSampled music sampling data set, from the website WhoSampled.com \citep{whosampled}, provided with permission.


\subsection{Temporal Networks}

The networks used in this evaluation are listed in Table \ref{tab:features}. So that the subgraph isomorphism task ran in a reasonable time, we selected a contiguous time-slice from each network such that the size was limited to \numprint{20000}. The networks include the start date but exclude the end date. We chose segments that represent the network in a somewhat steady state, without much growth or decline. The timespans covered by each extracted network, along with their order and size are also listed.

Our use of these networks, rather than the set of graphs used to test VF2 \citep{cordella-et-al-01} (randomly connected graphs, meshes, bounded valence graphs) is motivated by the type of application we envision for temporal analysis. In previous studies of cascades, the structures uncovered were predominantly composed of stars and long paths \citep{leskovec-06}, which are evident in many real networks.

\begin{table}[h]
\begin{center}
\begin{tabular}{l l r r r r }\hline
	Type & Network & Start & End & Order & Size \\ \hline  \hline
	\multirow{4}{*}{Communication}
		& Enron & Mar 2002 & Aug 2002 & \numprint{4729} & \numprint{18664} \\
		& Facebook & Sept 2004 & Aug 2006 & \numprint{5181} & \numprint{19169} \\
		& UC Irvine & Jun 2004 & Nov 2004 & \numprint{1251} & \numprint{16884} \\
		& Slashdot & Jun 2006 & Aug 2006 & \numprint{9937} & \numprint{19866} \\ \hline
	\multirow{1}{*}{Social}
		& Epinions & Jun 2003 & Jul 2003 & \numprint{2177} & \numprint{5288} \\ \hline
	\multirow{2}{*}{Time-respecting}
		& Hagelloch & 1861 & 1861 & \numprint{189} & \numprint{188} \\
		& WhoSampled & Jan 1984 & Jan 2001 & \numprint{21160} & \numprint{19986} \\ \hline
	\multirow{1}{*}{Transaction}
		& Prosper & Nov 2006 & Jan 2007 & \numprint{8690} & \numprint{72215} \\ \hline
\end{tabular}
\end{center}
\caption{The networks and their features.}
\label{tab:features}
\end{table}

In the set of communication networks we study, a node represents a user, and a directed edge represents a sent message (of which there can be many between any given pair of users). The Enron email network consists of emails sent between employees of Enron during the accounting scandal \citep{enron}. The Facebook network is drawn from users in the New Orleans region, and is composed of user-user wall posts \citep{facebook}. The UC Irvine network contains sent messages between the users of an online community of students from the University of California, Irvine \citep{opsahl}. The Slashdot data comes from the reply network of the technology website Slashdot, in which edges start from the replying user \citep{slashdot}. Within the Epinions social network, edges signify trust between users \citep{epinions}. Each user can decide whether to mark other users of the website as trusted or not. The Prosper Marketplace is an online peer-to-peer lending system, allowing borrowers and lenders to interact without bank intermediation \citep{prosperwebsite}. Edges are directed from lender to borrower, representing the flow of funds in the network.

We also explore two networks which are time-respecting by construction (given an infinite $d$-value). This happens because no path can exist that is not time-respecting. The Hagelloch epidemic data represent a measles outbreak in Hagelloch, Germany in 1861 \citep{hagelloch}. In this case, individuals were only infected once. Thus, the case where individual $v$ infects $w$ and then $u$ infects $v$ cannot occur, since individual $v$ was already infected. The WhoSampled network has a similar property \citep{whosampled}. In this network, an edge is directed from the sampled song to the song containing the sample. So, the case in which song $w$ samples $v$ and then $v$ samples $u$ cannot occur, since song $v$ was already created.


\subsection{Query Graphs}

\begin{figure}[]
	\centering
	
	\subfloat[]
		{\label{fig:q1}\includegraphics[width=0.26\textwidth, height=0.1\textheight]{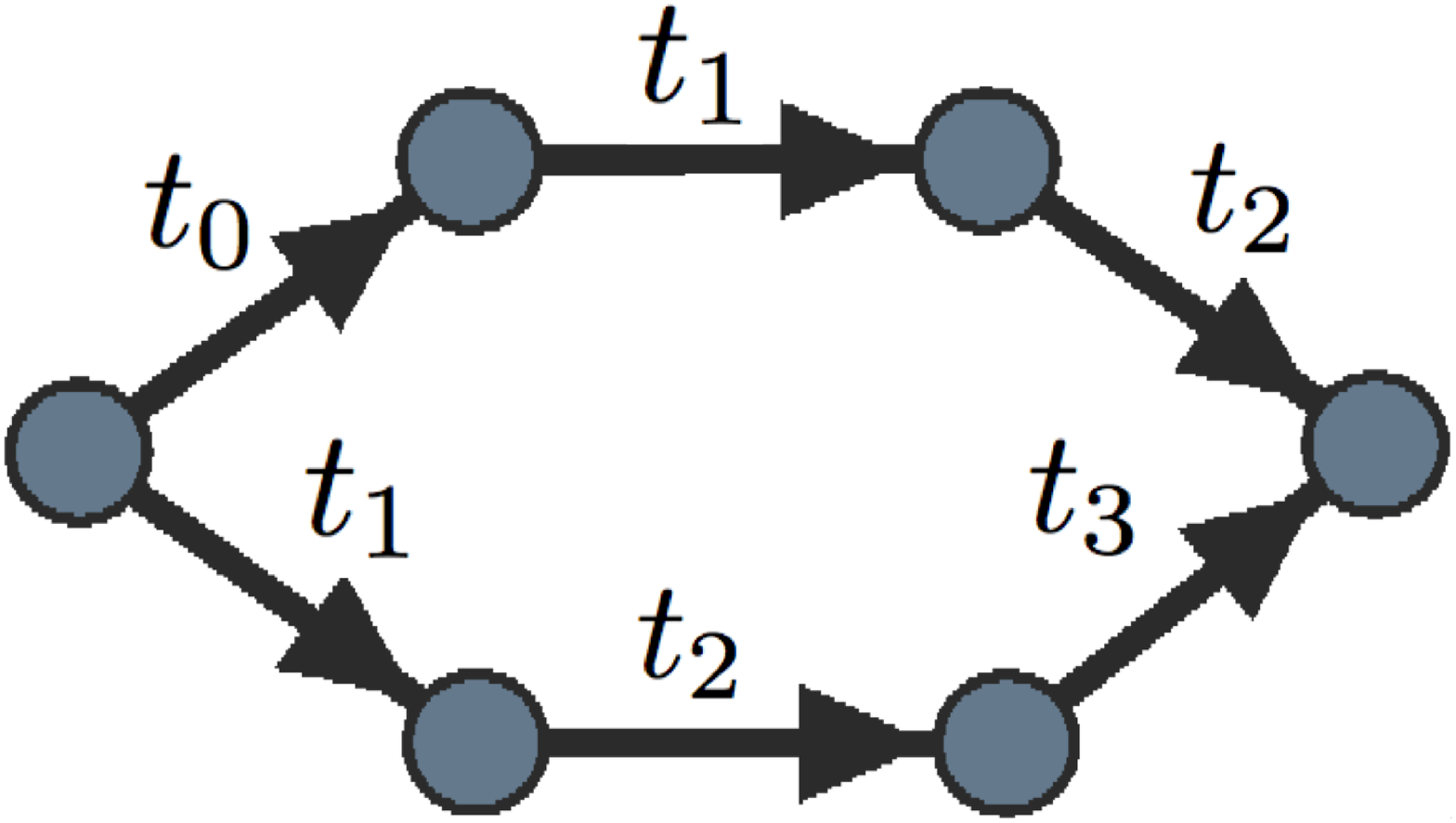}}
	\quad
	\subfloat[]
		{\label{fig:q2}\includegraphics[width=0.19\textwidth, height=0.07\textheight]{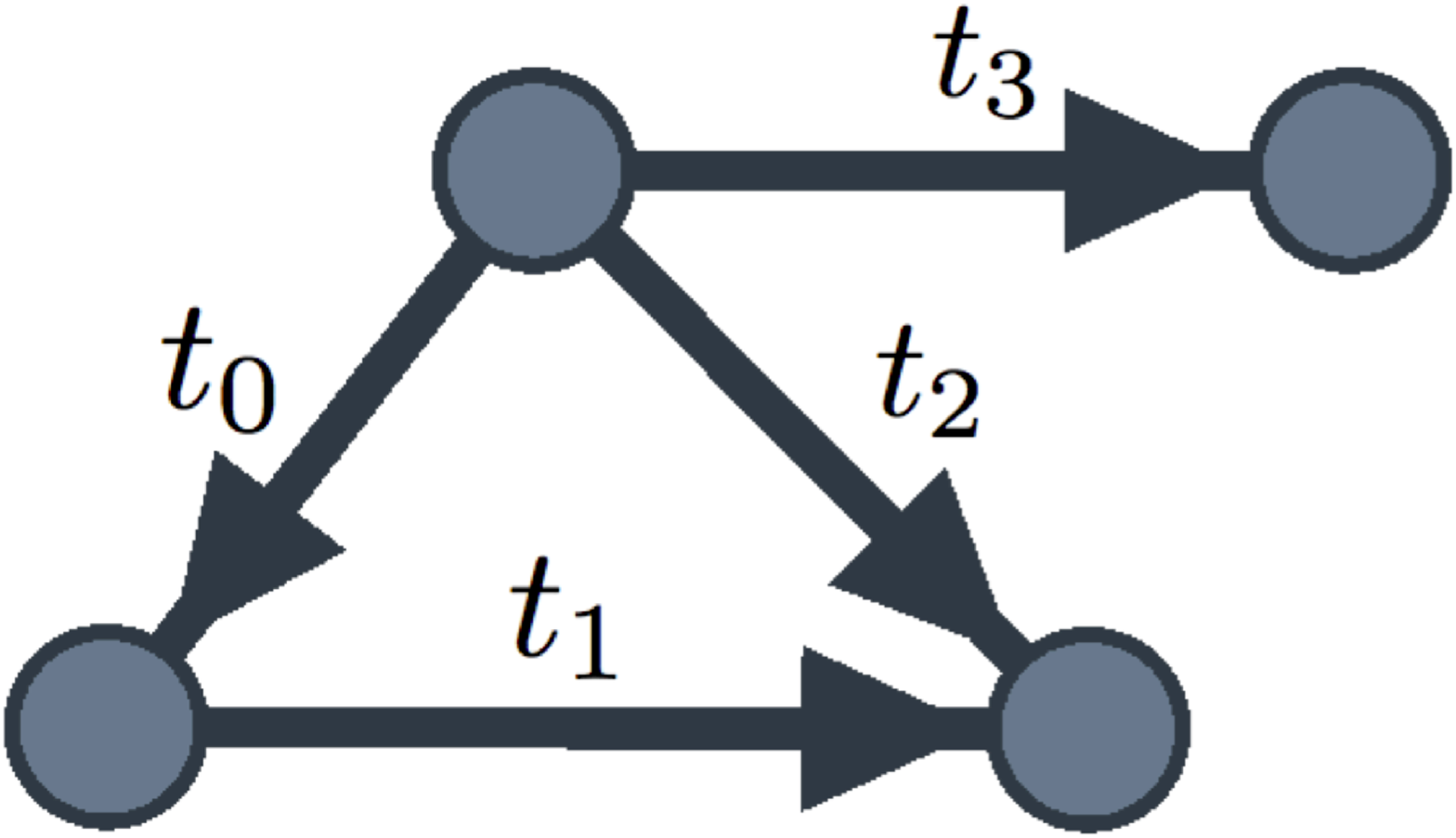}}
	\quad
	\subfloat[]
		{\label{fig:q3}\includegraphics[width=0.23\textwidth, height=0.07\textheight]{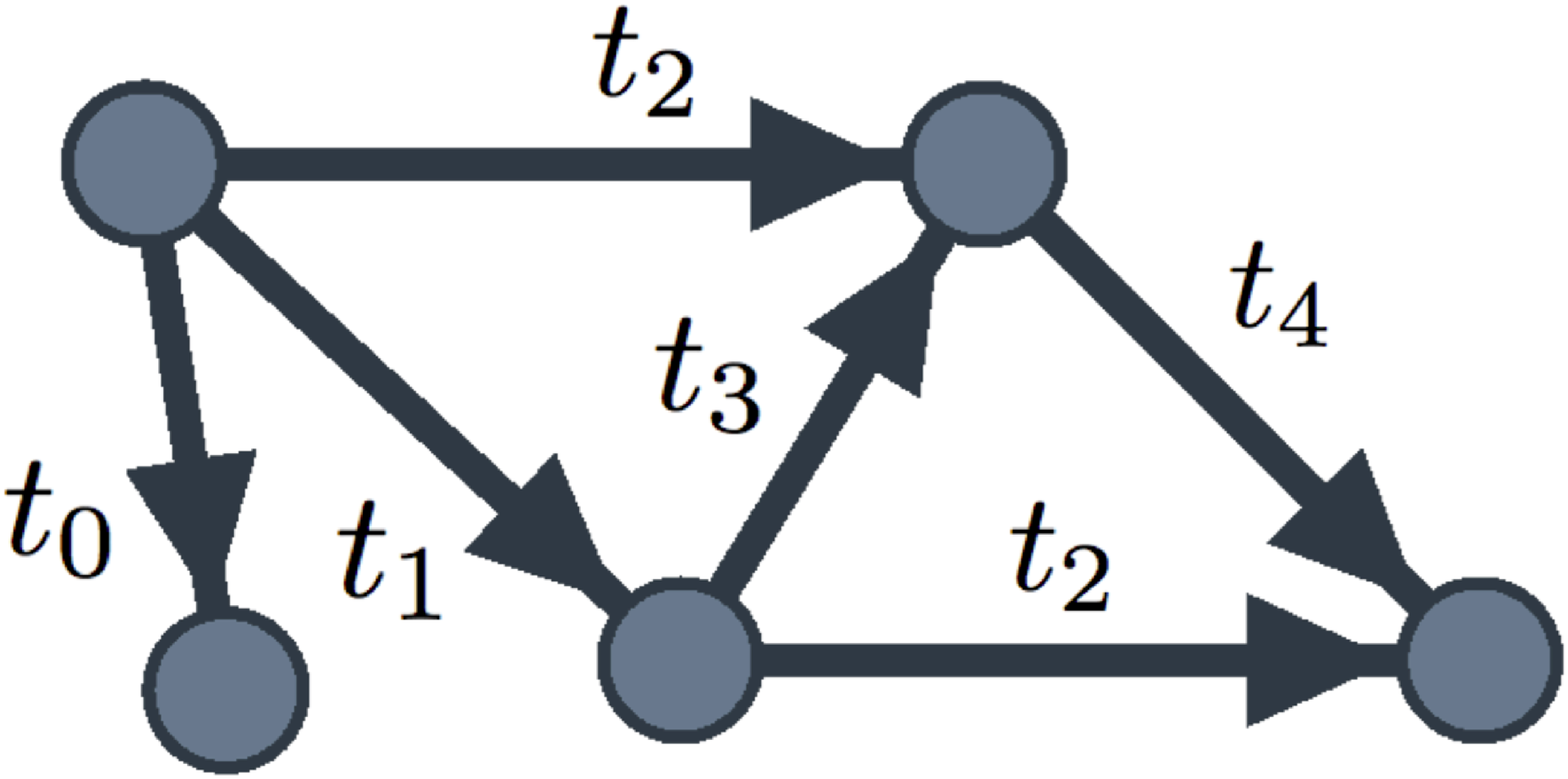}}
	\quad
	\subfloat[]
		{\label{fig:q4}\includegraphics[width=0.24\textwidth, height=0.11\textheight]{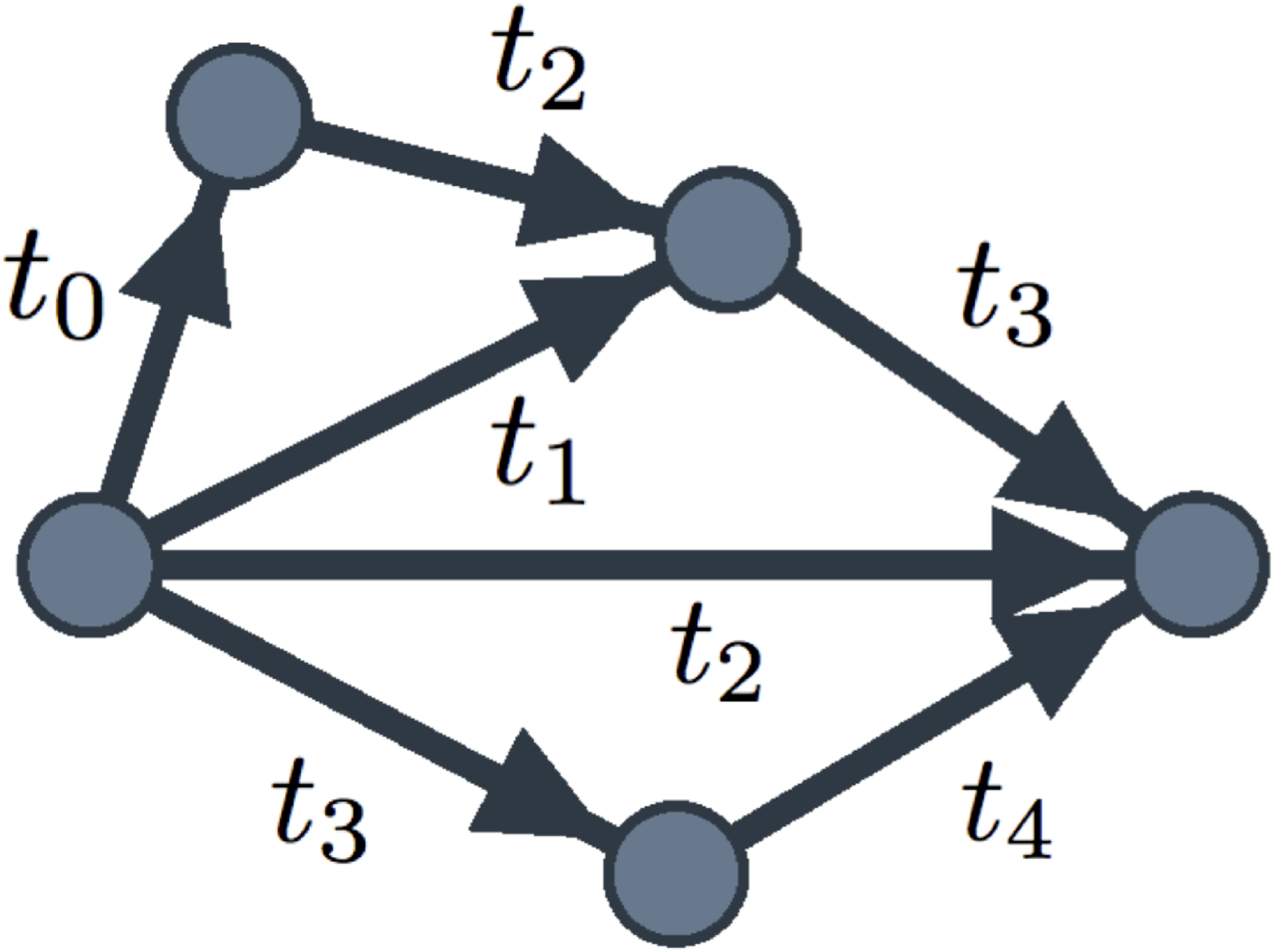}}	
		
	\caption{A sample set of query graphs from the 30 we use in our evaluation. It is the embeddings of these topologies in the network data that must be time-respecting. The time-stamp subscripts, which show one possible permissible ordering of interactions, convey this abstraction. Permissible values of the time-stamps must ensure that the embeddings are time-respecting.}
	\label{fig:q}
\end{figure}

When we specify a given query graph, its topology dictates the topology of the embedding we aim to examine in the network. It is the pattern embedded in the network that must be time-respecting. The temporal aspect of the query graphs, just like their topology, is user-defined. Since we do not constrain the search to sets of interactions that took place at specific times, we reflect this in our construction of the query graphs, by not putting explicit time-stamps on the interactions. It is the time-stamps on interactions in the network data that are relevant. We emphasize this in \figurename{\ref{fig:q}} by using abstract time-stamps, which illustrate our requirement that embeddings of the query graphs must be time-respecting, for any given values of the time-stamps.

The query graphs comprise 30 small graphs. The set is chosen somewhat arbitrarily, aiming to have an even distribution of graph sizes, with between three and eight nodes. Their small size enables simpler interpretation of embeddings in the data. A set of examples is shown in \figurename{\ref{fig:q}}. Our choice of query graphs does not consider the topology of the networks under study. Some query graphs do not appear in the networks we analyzed, and some have many embeddings.

\figurename{\ref{fig:q1}} is a fan-out-fan-in graph, with two paths and three hops. If this represented communication -- for example in the Facebook network -- the left-most individual would initiate the communication, which would be forwarded through intermediaries, who would jointly pass the message to a final individual. If the context was the WhoSampled network, the left-most node would represent a song, which is sampled twice. The sampling continues until the right-most song uses samples from the two songs to its left. In a peer-to-peer transaction network, the node on the left would lend money to a pair of people, who would lend this money to others, who eventually lend to the right-most person. Such a pattern may indicate suspicious behaviour, such as money laundering or arbitrage. 

Considering the query in \figurename{\ref{fig:q3}}, in the context of the Epinions network, we might interpret it as follows. The top-leftmost user could mark the bottom-leftmost user as trusted. Then, this trusted user could mark the bottom-rightmost user as not trusted. Trusting that example, the top-leftmost user might agree and mark that individual as not trusted. Then the top-leftmost user could mark someone else as trusted.

Taking the example in \figurename{\ref{fig:q3}}, we can interpret its embedding in the context of the temporal networks. The query pattern comprises some well-known network motifs \citep{milo-02} -- it begins with a fan-out motif, running into a feed-forward motif. In the Enron network, the leftmost individual would email three other employees. Two of these, along with another employee, would then form a feed-forward loop in emailing another employee.

The example in \figurename{\ref{fig:q4}} contains a mix of a feed-forward and fan-out-fan-in motif. It can be interpreted in a manner similar to that described above. The meaningfulness of our model resides in the fact that interactions take place close in time, and there is the potential for flow among network entities.

\section{Results}\label{sec:results}

Our proposed algorithms were implemented in the programming language Python \citep{python}, using the NetworkX library \citep{networkx}. The VF2 algorithm is included in this library, implemented as part of a project at the Complexity Sciences Center and Physics Department, UC Davis. We further developed this implementation to process temporal networks and use temporal information within the matching process. Our extended implementation can handle directed graphs as well as directed multigraphs (graphs with multiple interactions between nodes).

It is important to note that, no matter which of the three approaches are used, the same set of embeddings will be found. This is emphasized in \figurename{\ref{fig:schematic_time}} and \figurename{\ref{fig:schematic_top}}. Each algorithm produces accurate solutions. The difference is in the amount of time taken for the algorithms to run. Thus, our evaluation deals only with the execution time, in seconds, of the algorithms. 

The extraction of time-respecting subgraphs proved the most computationally expensive method. Since there may be many internally consistent time-respecting subgraphs which overlap to a large extent, the recursive search for these subgraphs can be very extensive. In practice, the algorithm exceeded the maximum recursion depth (on a Linux server limited to 5GB of physical memory), even for some of our smallest networks. Thus, we restrict our evaluation to {\tempvf} and {\vf}.

In this section, we motivate our choice of $d$-value for each network, which depends on the time-scale at which interactions take place within the network. We note the time taken for the computations, and compare the speeds of our two best approaches, {\tempvf} and {\vf}. We then discuss some of the factors which may have an impact on the differences in execution time.


\subsection{Choosing the $d$-value}

\begin{figure}[]
	\centering
	\subfloat[Enron]
		{\label{fig:d_enron}\includegraphics[width=0.45\textwidth, height=0.21\textheight]{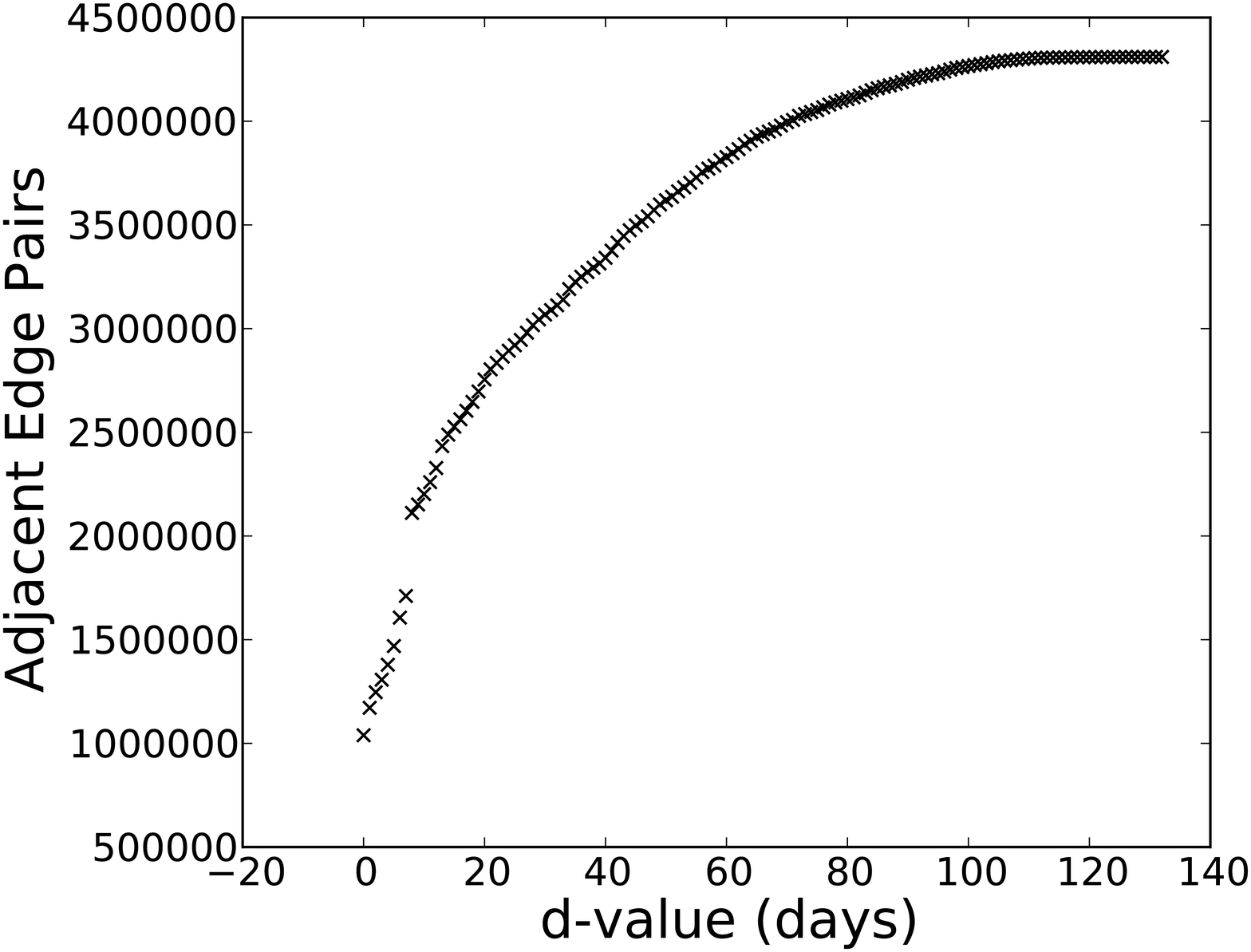}}
		\qquad
	\subfloat[Facebook]
		{\label{fig:d_facebook}\includegraphics[width=0.45\textwidth, height=0.21\textheight]{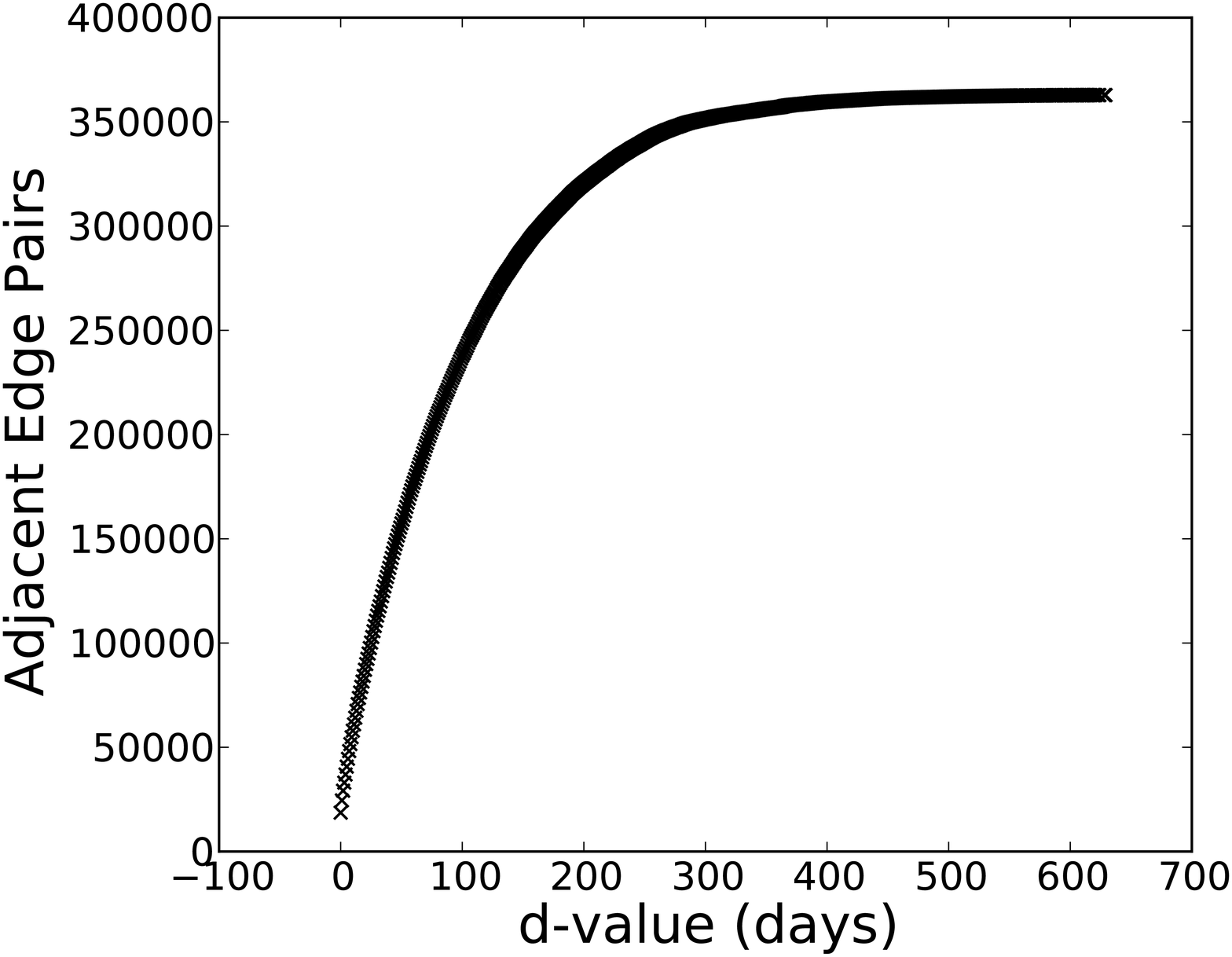}}\\
		
	\subfloat[UC Irvine]
		{\label{fig:d_opsahl}\includegraphics[width=0.45\textwidth, height=0.21\textheight]{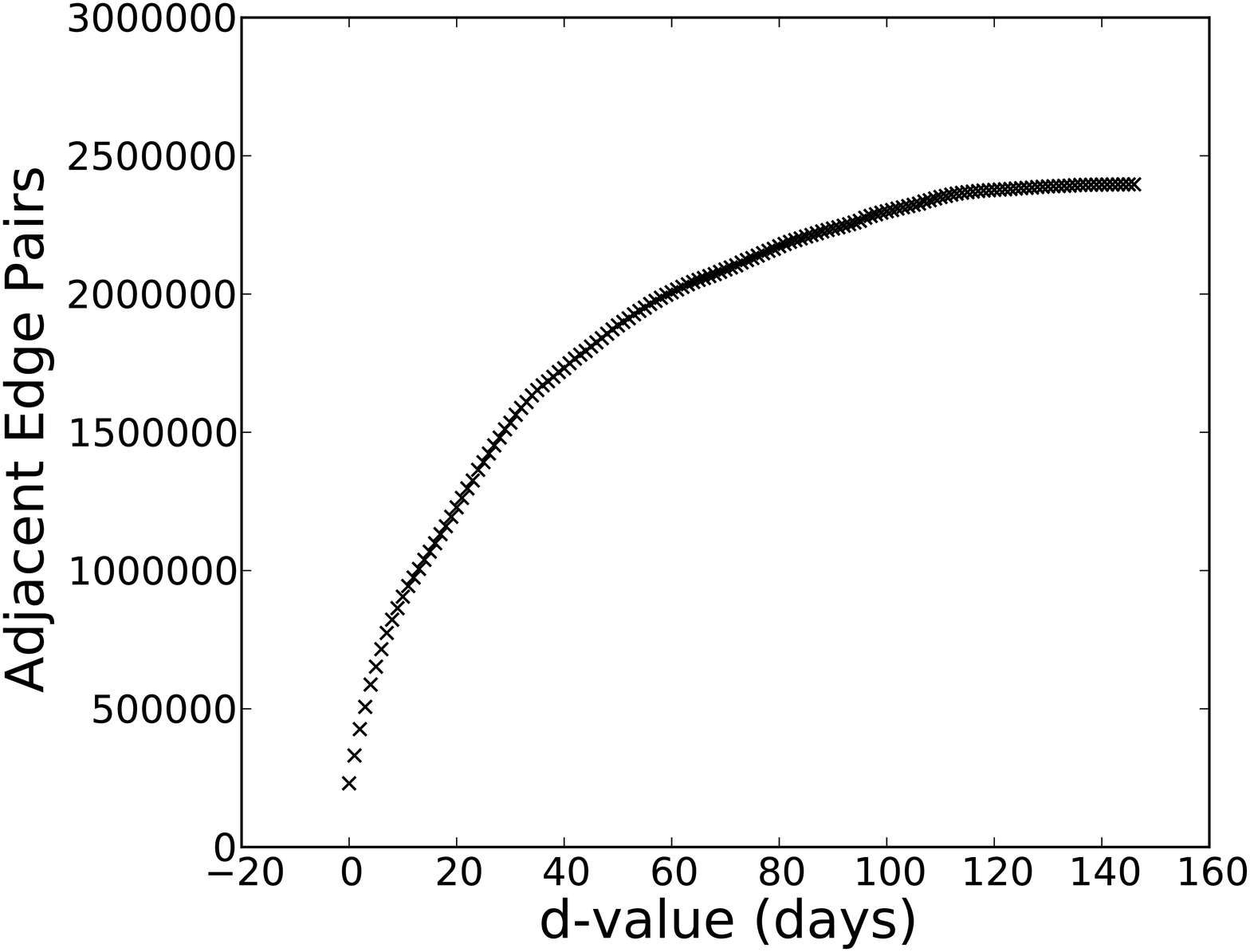}}
		\qquad
	\subfloat[Slashdot]
		{\label{fig:d_slashdot}\includegraphics[width=0.45\textwidth, height=0.21\textheight]{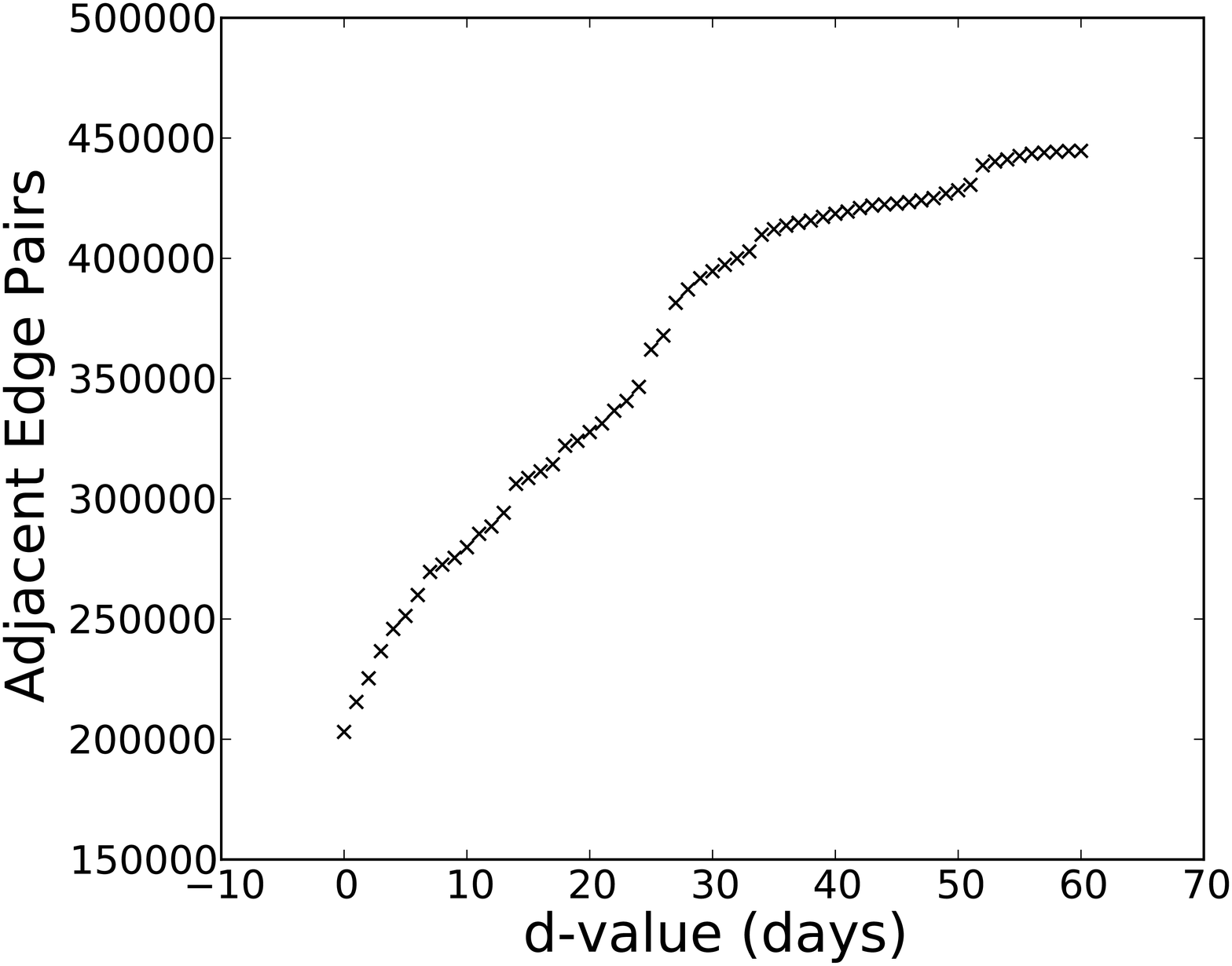}}\\
		
	\subfloat[WhoSampled]
		{\label{fig:d_who}\includegraphics[width=0.45\textwidth, height=0.21\textheight]{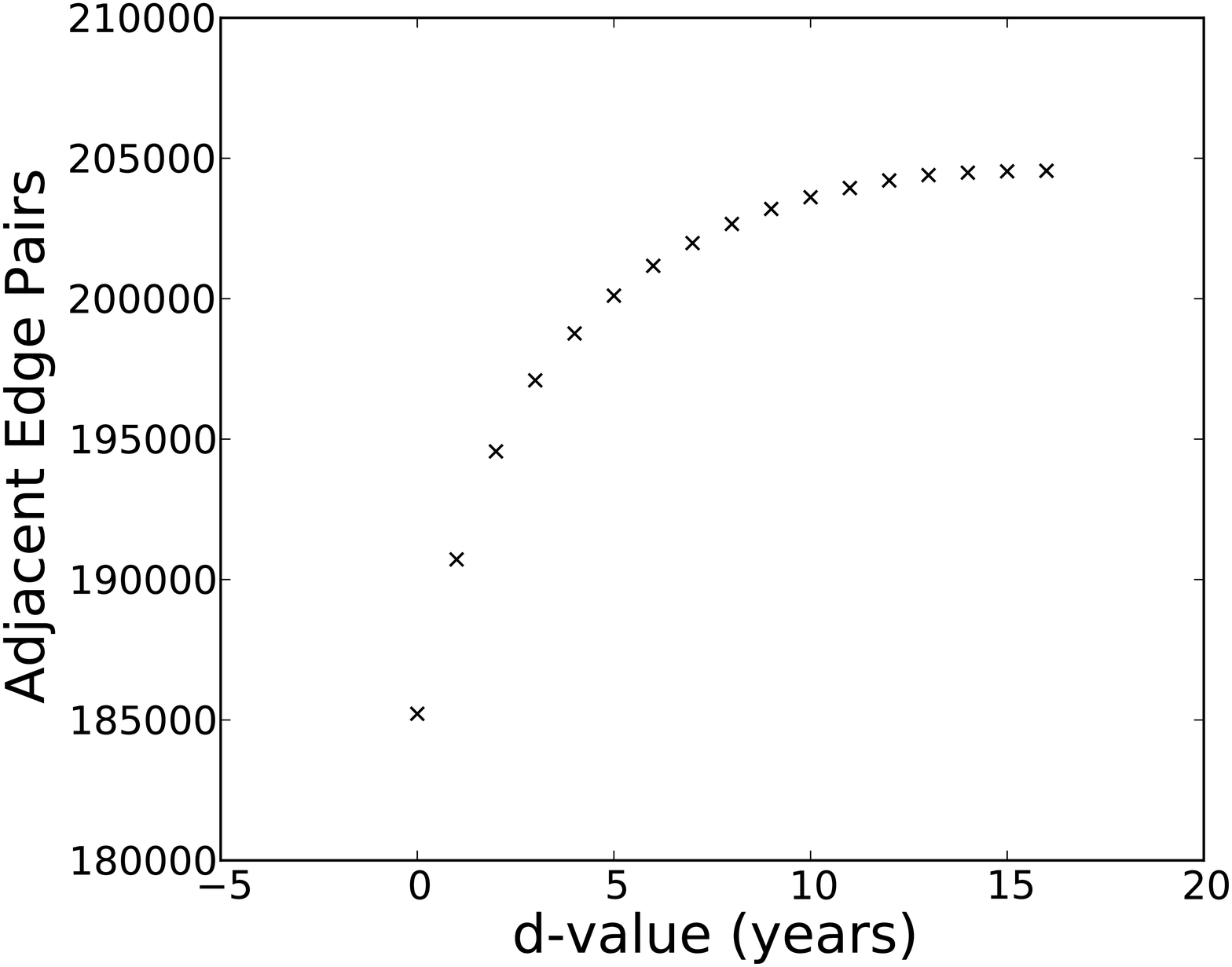}}
		\qquad
	\subfloat[Hagelloch]
		{\label{fig:d_hagelloch}\includegraphics[width=0.45\textwidth, height=0.21\textheight]{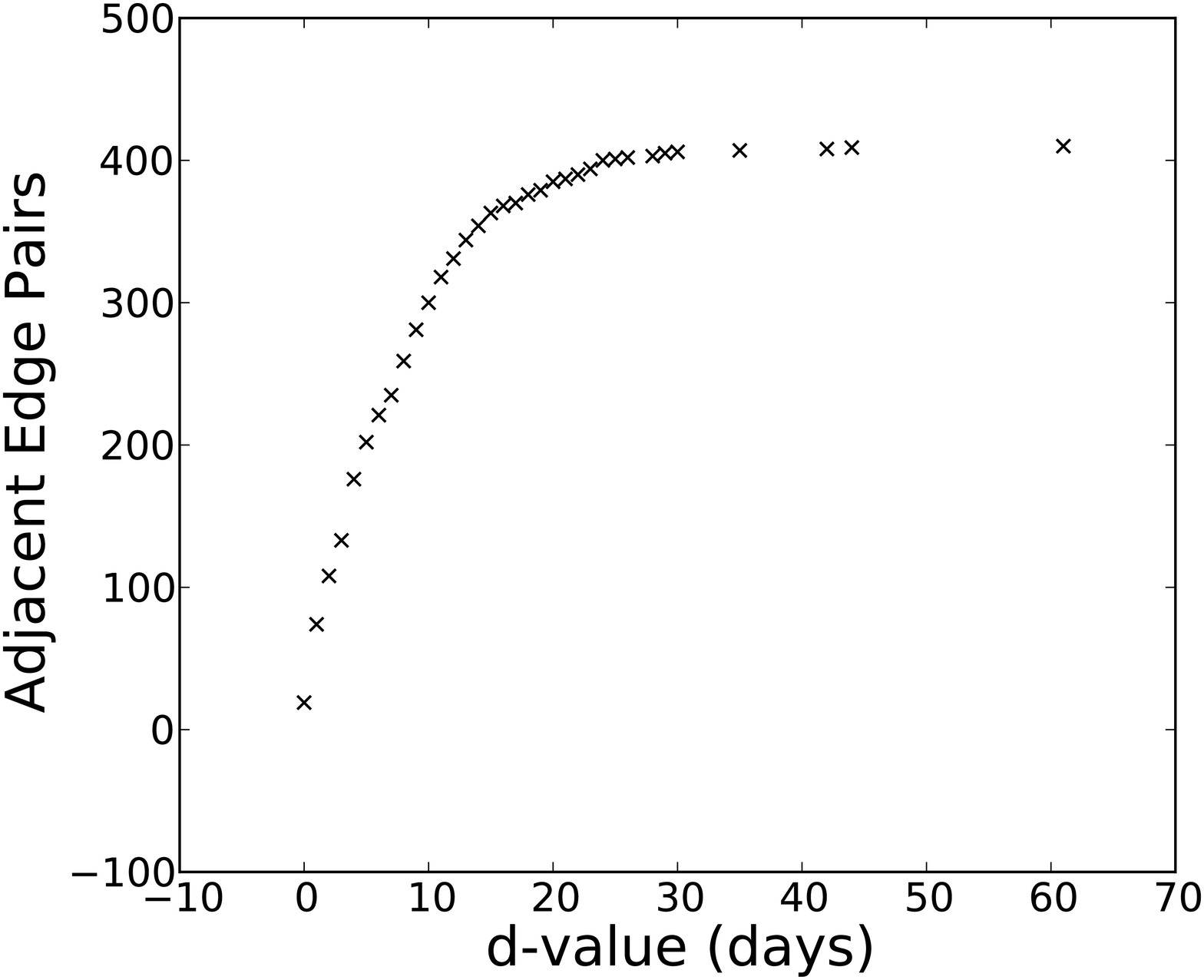}}\\
		
	\subfloat[Prosper]
		{\label{fig:d_prosper}\includegraphics[width=0.45\textwidth, height=0.21\textheight]{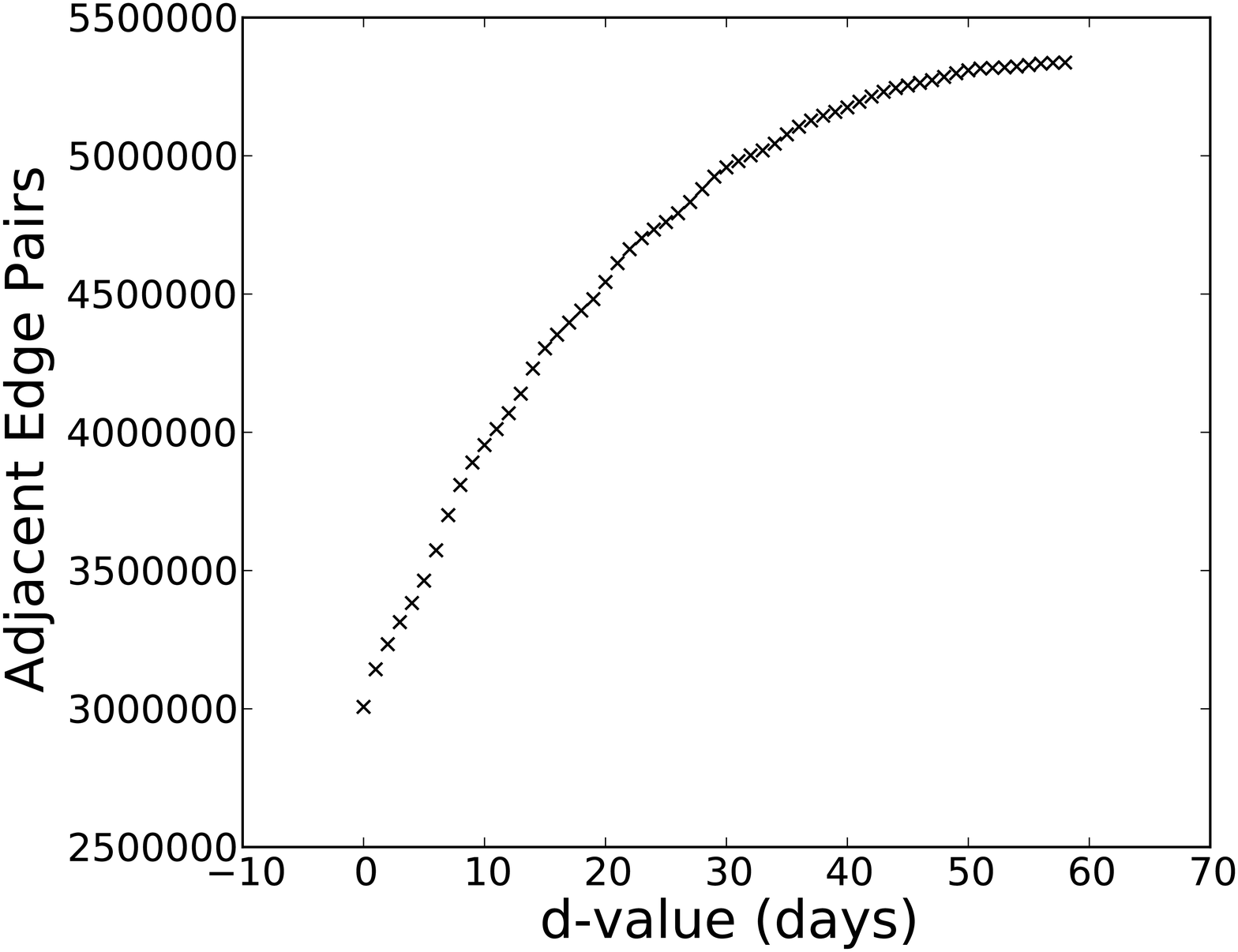}}
		\qquad
	\subfloat[Epinions]
		{\label{fig:d_epinions}\includegraphics[width=0.45\textwidth, height=0.21\textheight]{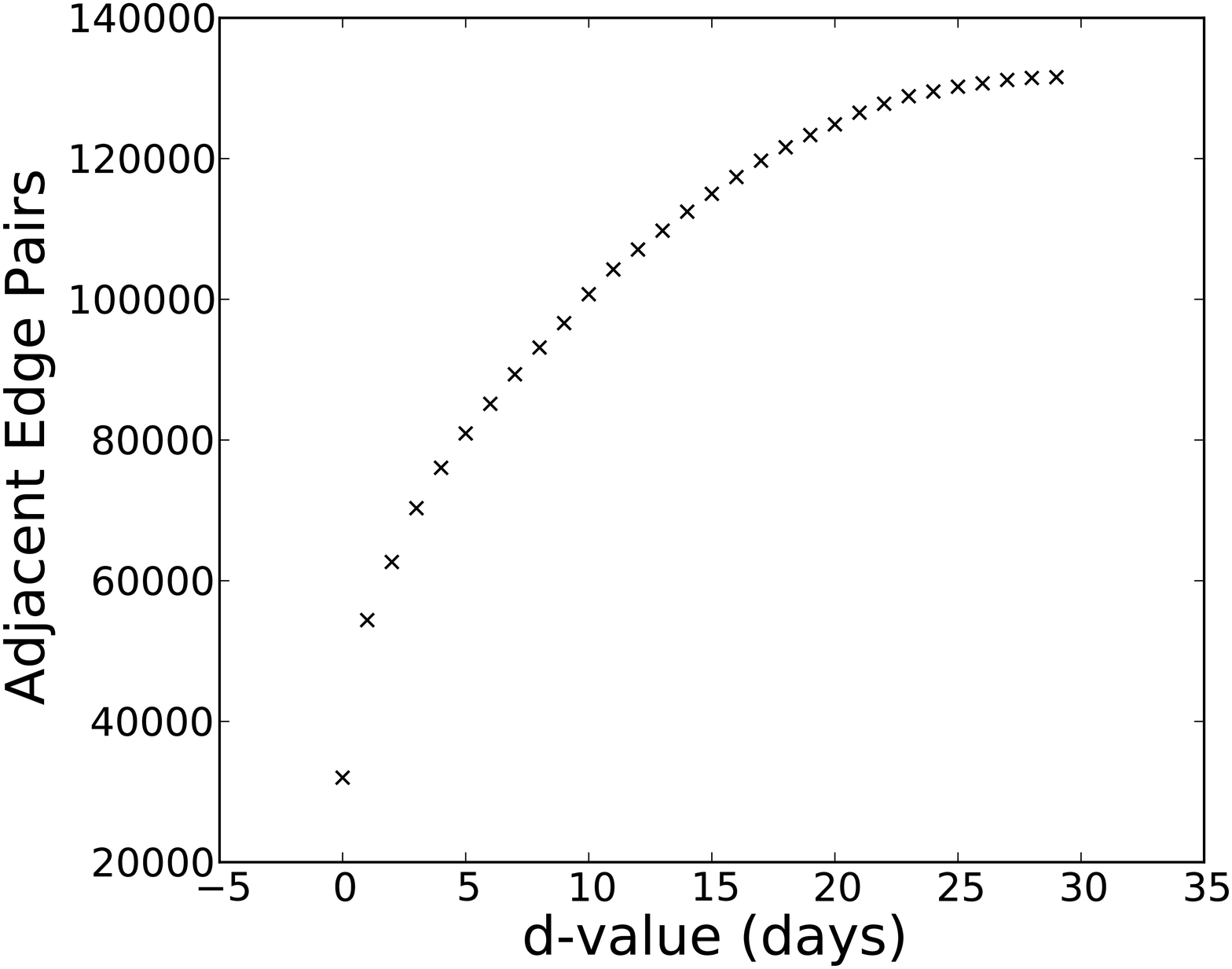}}\\
		
	\caption{The $d$-value distributions of the network data sets used in our evaluation. The x-axis shows the time-scale at which interactions between entities took place. The y-axis shows a cumulative count of how many times such an interaction took place within the amount of time specified by the x-axis.}
	\label{fig:d_dist}
\end{figure}

The choice of $d$-value must make sense for each application, to ensure meaningful results. For example, in an email network, emails might be forwarded or replied to over a number of days, whereas in an online social network, interactions might take place over the course of minutes or hours. To elucidate the range of $d$-values that might occur in a network, we compare the time-stamps on every adjacent pair of interactions. Plots of the results are shown in \figurename{\ref{fig:d_dist}}. If the time allowed between interactions is one hour, then a pair of interactions with five minutes between them certainly qualifies. Thus, the plots are cumulative.

In \figurename{\ref{fig:d_facebook}} for example, the plot visibly plateaus after a certain point. We posit that the few larger values may be ignored for our purposes. We take this elbow point to be the largest meaningful $d$-value. The values we choose for our evaluation are 10\%, 20\%, 30\%, 40\% and 50\% of the largest accepted value. This allows us to evaluate the pruning power of $d$ over a relevant time-scale. The precise $d$-values for each network are listed in Table \ref{tab:d_value}. The music sampling network is evaluated over a period of years, since musicians sometimes sample much older as well as more recent music. A scale of days is used for the communication and social networks, reflecting how the users interacted via social media. The disease network also occurs over a number of days.

\begin{table}[]
	\begin{center}
		\begin{tabular}{lllr} \hline
			Time Unit & Network & $d$-values & $d\_max$ \\ \hline  \hline
			\multirow{6}{*}{Days} 
				& Enron & 10, 20, 30, 40, 50 & 100 \\
				& Epinions & 2, 4, 6, 8, 10 & 20 \\
				& Facebook & 30, 60, 90, 120, 150 & 300 \\
				& Hagelloch & 2, 4, 6, 8, 10 & 20 \\
				& Prosper & 2, 4, 6, 8, 10 & 20 \\
				& Slashdot & 3, 6, 9, 12, 15 & 30 \\
				& UC Irvine & 10, 20, 30, 40, 50 & 100 \\ \hline
			\multirow{1}{*}{Years}
				& WhoSampled & 1, 2, 3, 4, 5 & 10 \\ \hline
		\end{tabular}
	\end{center}
	\caption{The interactions within each network occur over a given time-scale. The music sampling network has the longest inter-contact duration. The maximum $d$-value for each network ($d\_max$) was taken from the elbow of each $d$-value distribution plot, shown in \figurename{\ref{fig:d_dist}}. The $d$-values we use in our evaluation occur at 10\%, 20\%, 30\%, 40\% and 50\% of $d\_max$ for each network.}
	\label{tab:d_value}
\end{table}


\subsection{Processing Time}

\begin{figure}[]
	\centering
	\subfloat[Enron ($d$-value 30 days)]
		{\label{fig:raw_enron}\includegraphics[width=0.43\textwidth, height=0.21\textheight]{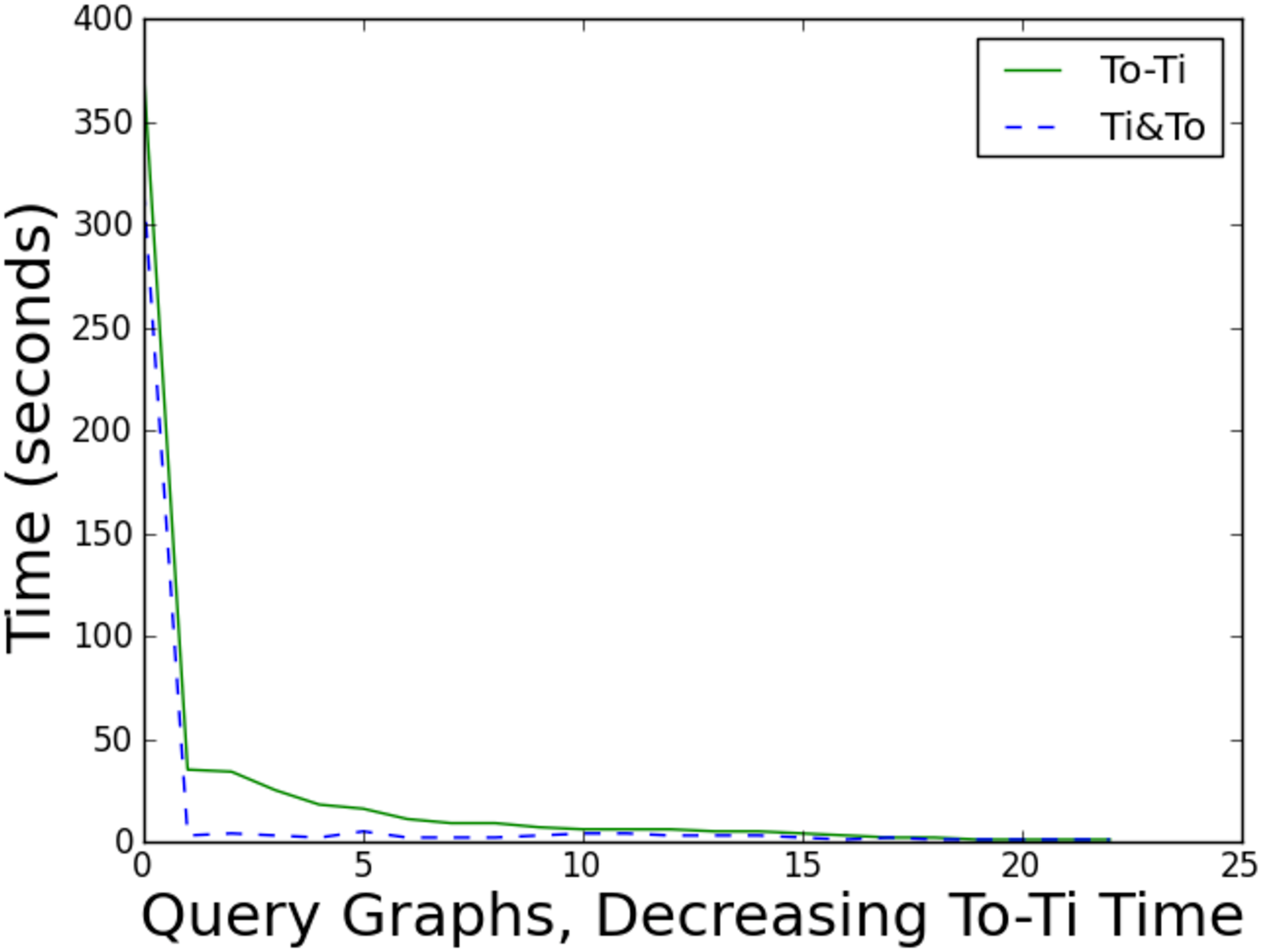}}
		\qquad
	\subfloat[Facebook ($d$-value 90 days)]
		{\label{fig:raw_facebook}\includegraphics[width=0.43\textwidth, height=0.21\textheight]{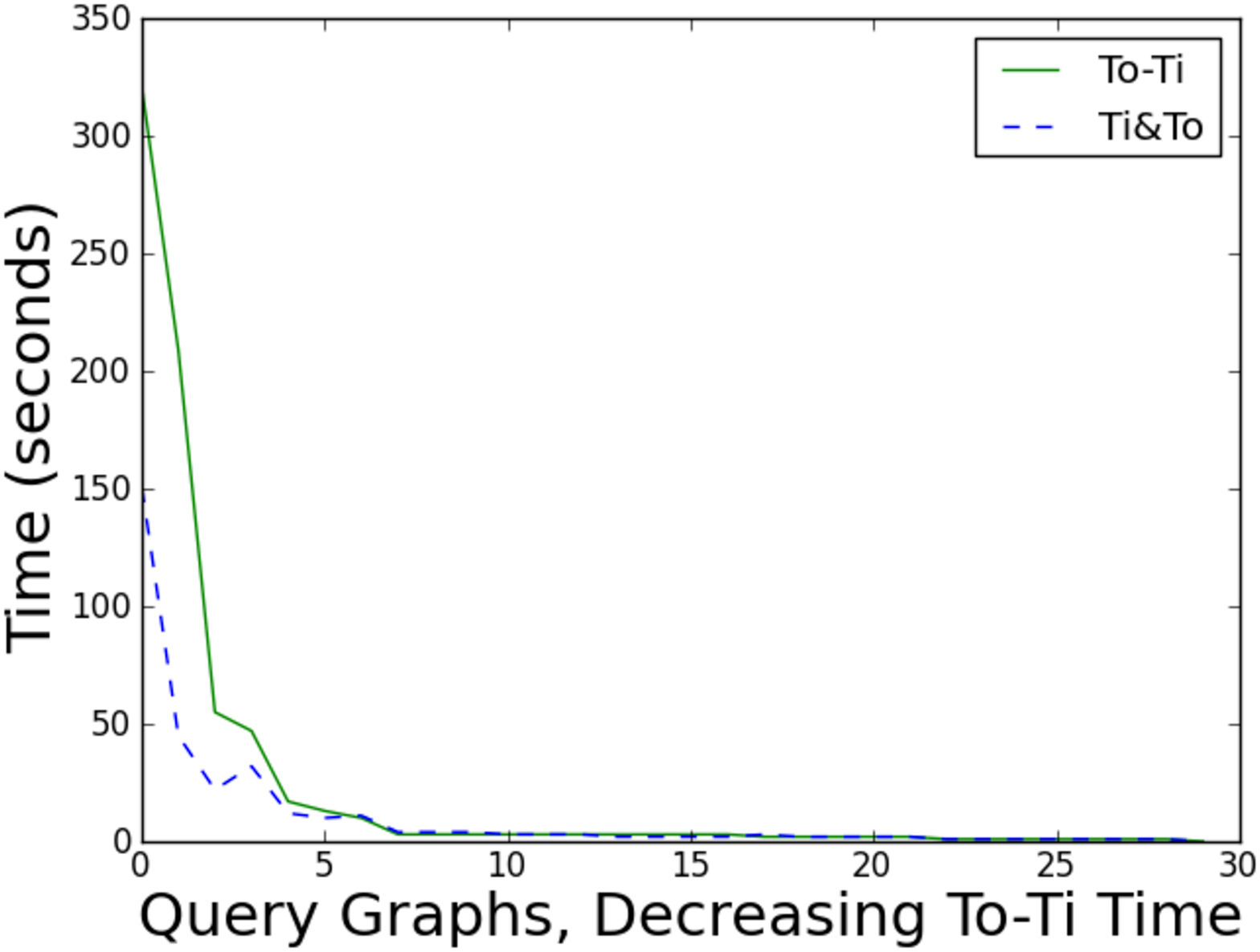}}\\
	
	\subfloat[UC Irvine ($d$-value 30 days)]
		{\label{fig:raw_opsahl}\includegraphics[width=0.43\textwidth, height=0.21\textheight]{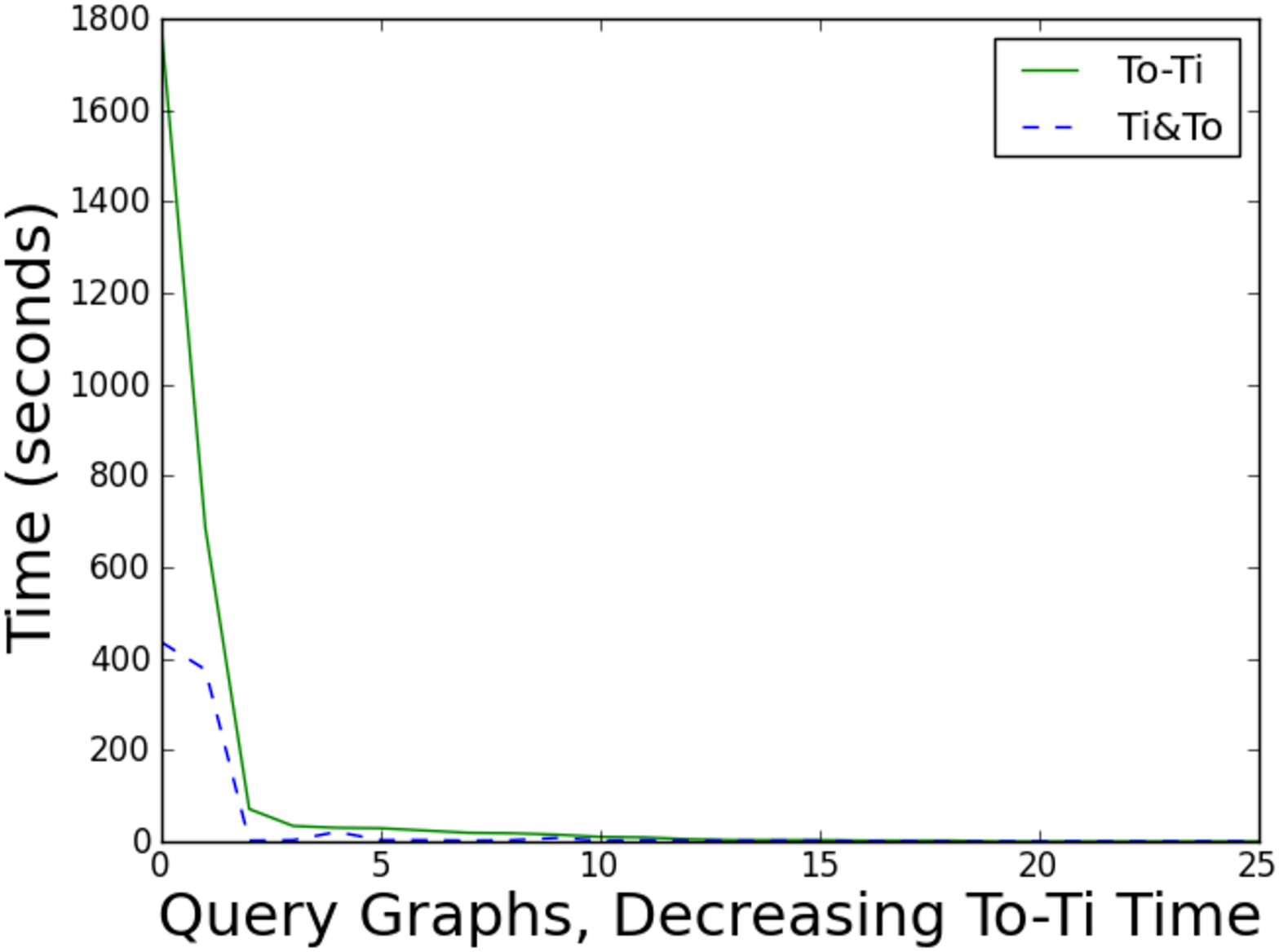}}
		\qquad
	\subfloat[Slashdot ($d$-value 9 days)]
		{\label{fig:raw_slashdot}\includegraphics[width=0.43\textwidth, height=0.21\textheight]{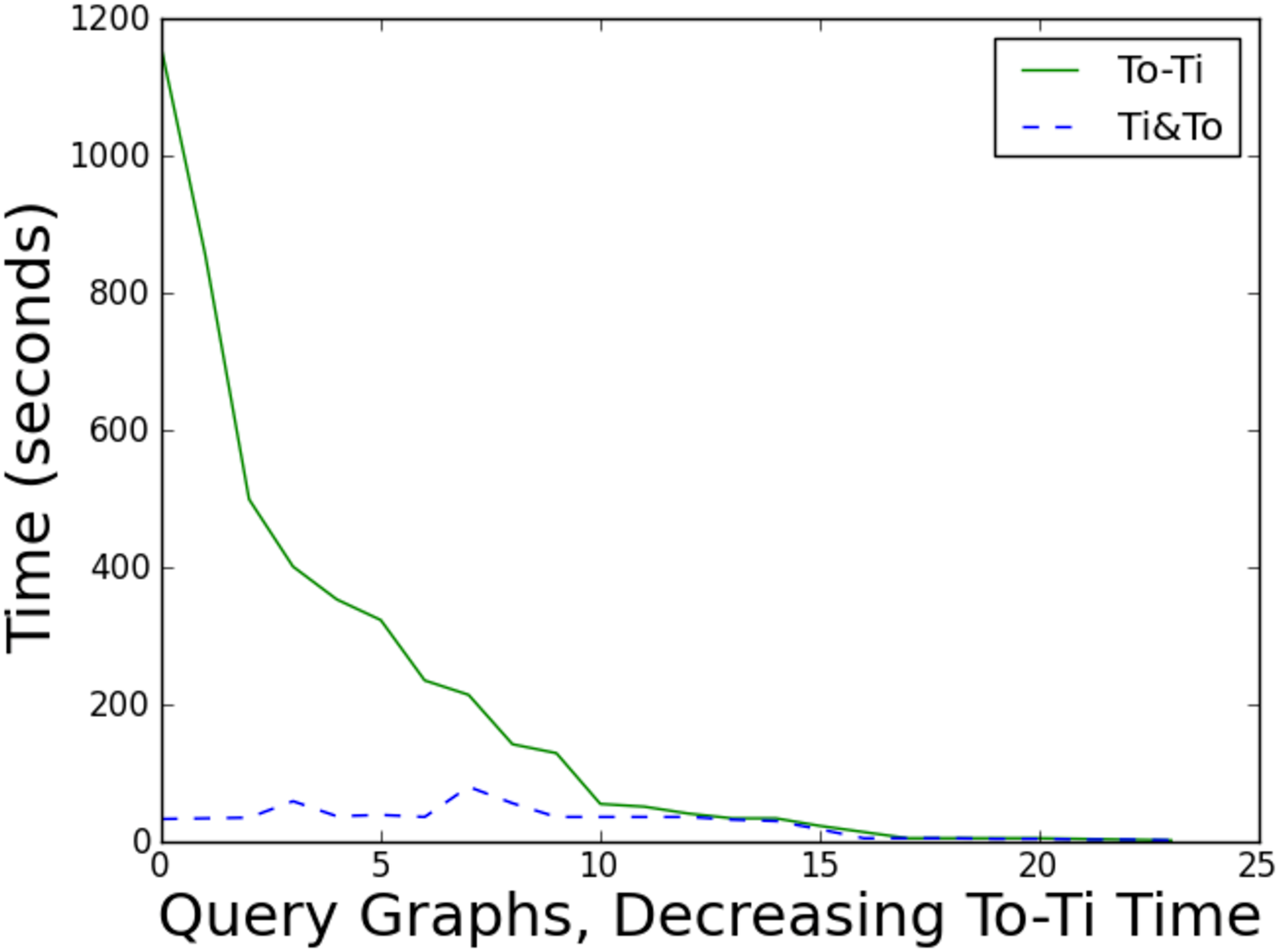}}\\
	
	\subfloat[WhoSampled ($d$-value 3 years)]
		{\label{fig:raw_who}\includegraphics[width=0.43\textwidth, height=0.21\textheight]{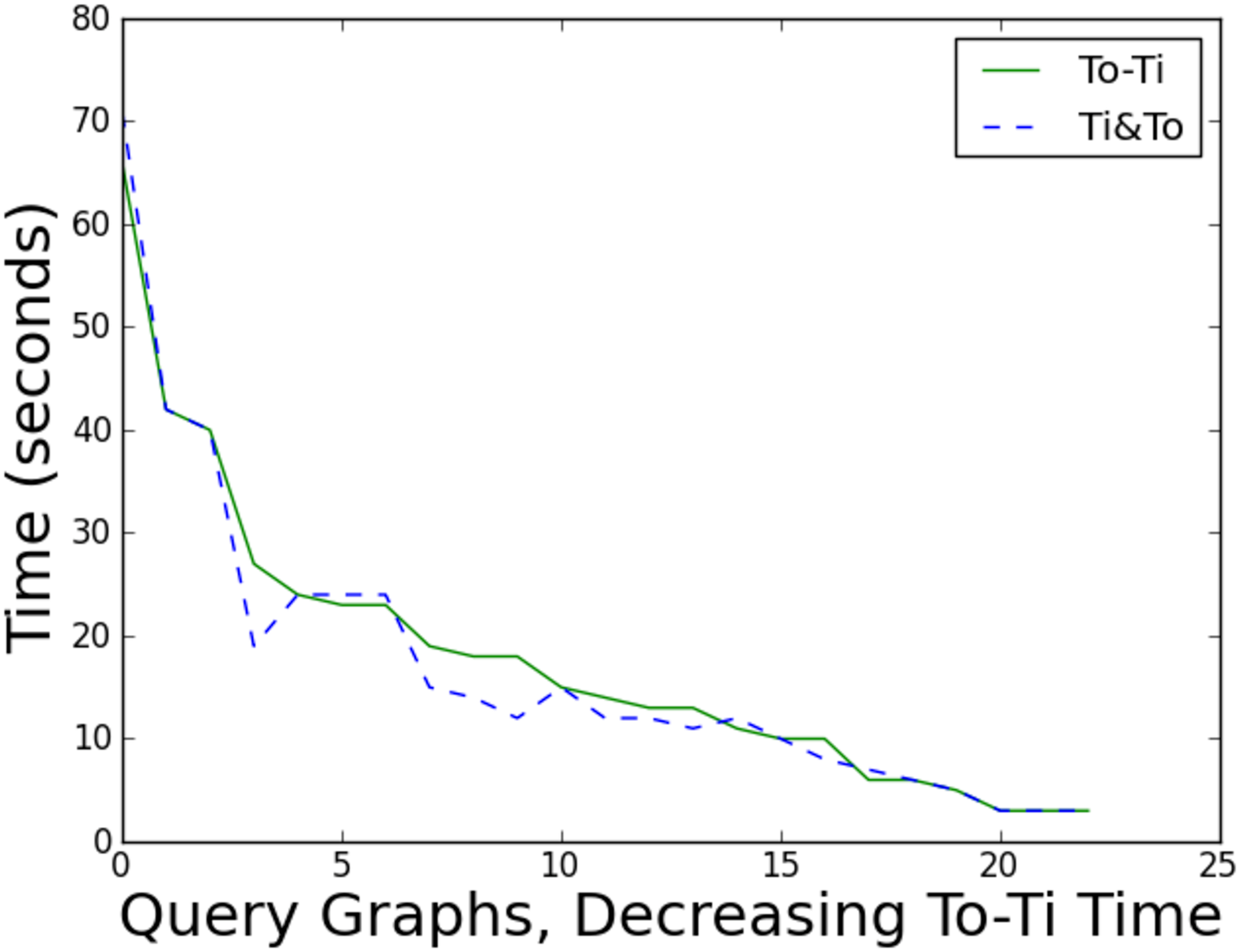}}
		\qquad
	\subfloat[Hagelloch ($d$-value 6 days)]
		{\label{fig:raw_hagelloch}\includegraphics[width=0.43\textwidth, height=0.21\textheight]{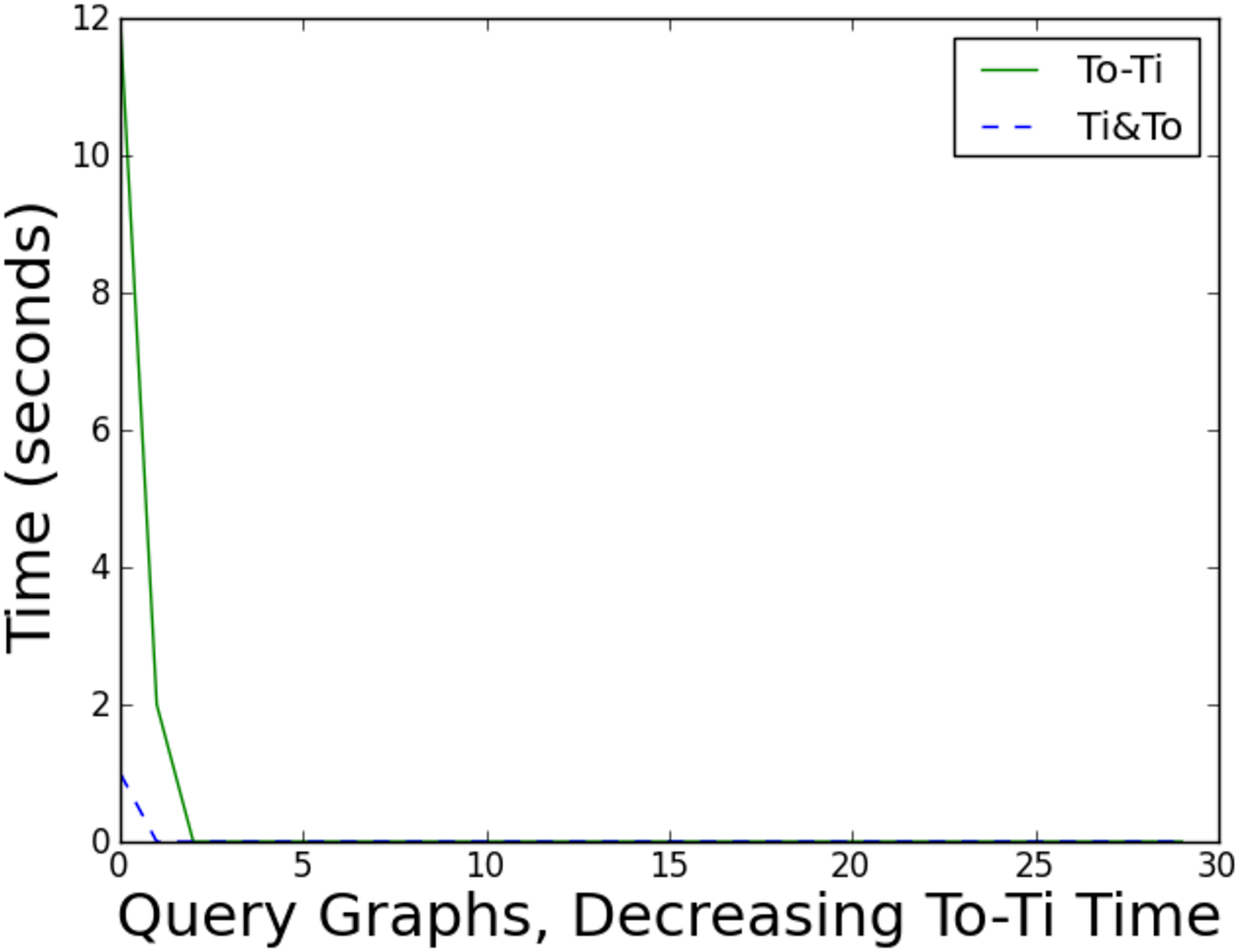}}\\
	
	\subfloat[Prosper ($d$-value 6 days)]
		{\label{fig:raw_prosper}\includegraphics[width=0.43\textwidth, height=0.21\textheight]{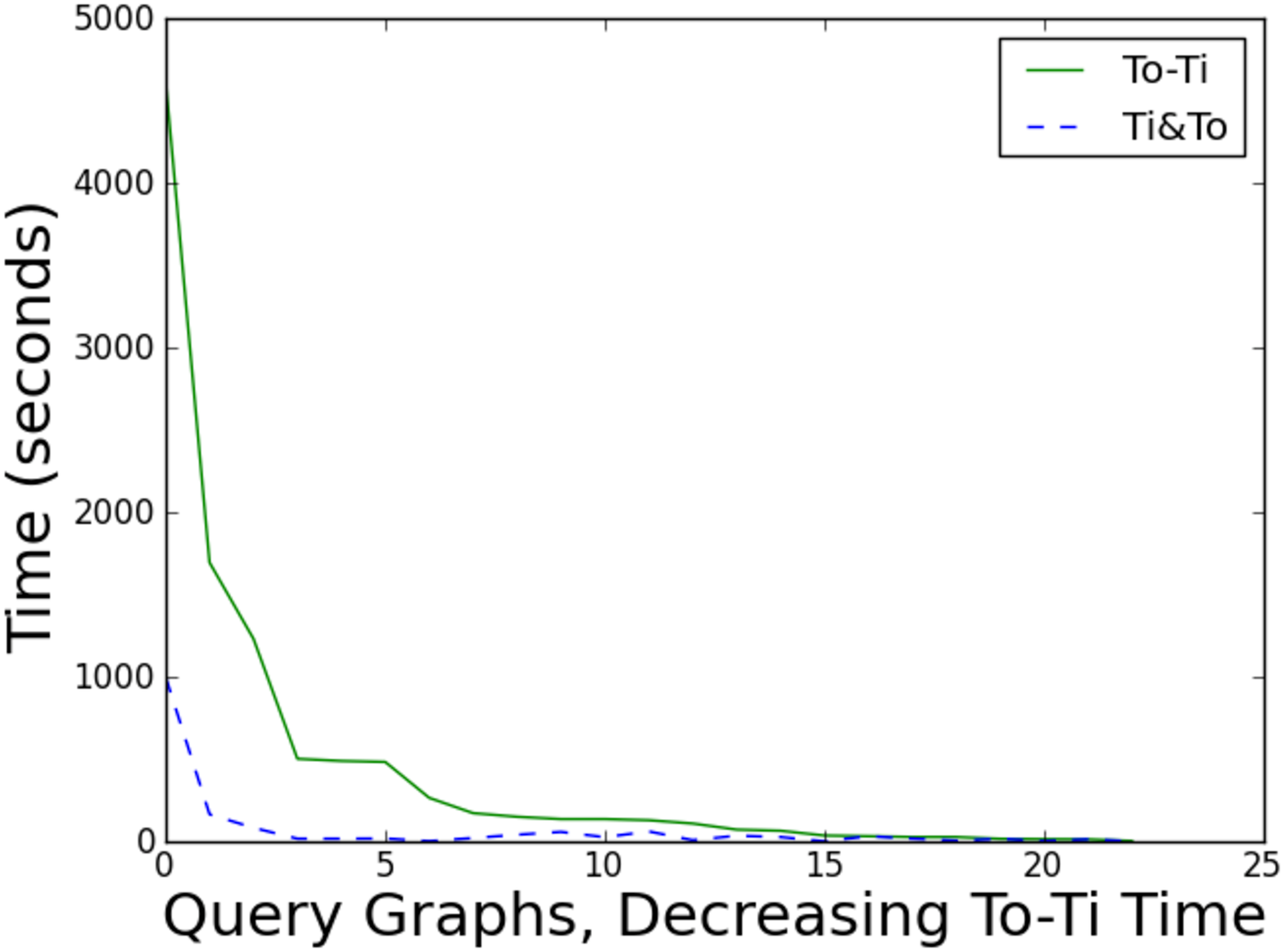}}
		\qquad
	\subfloat[Epinions ($d$-value 6 days)]
		{\label{fig:raw_epinions}\includegraphics[width=0.43\textwidth, height=0.21\textheight]{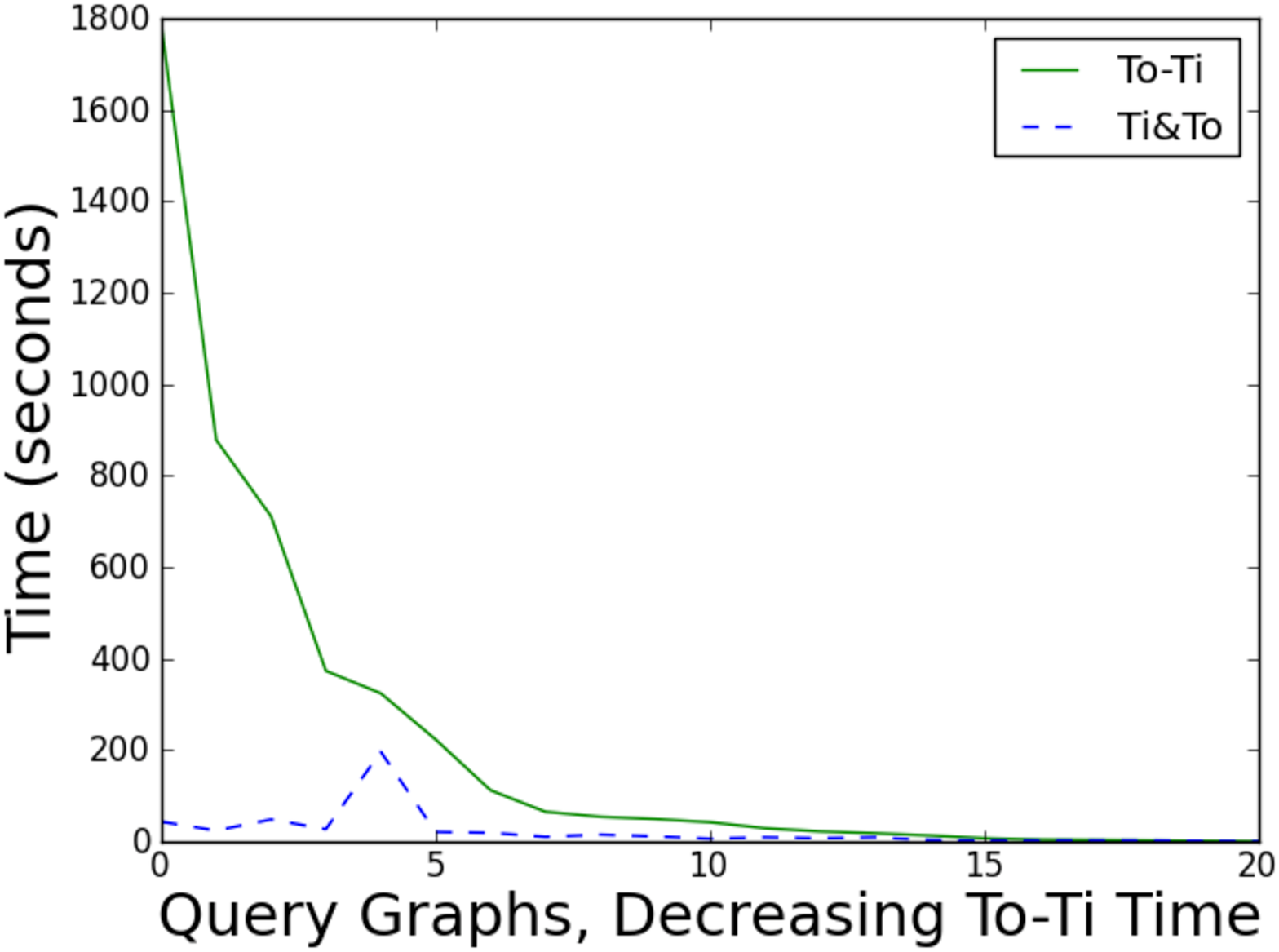}}\\

	\caption{Comparing the time taken (in seconds) for {\tempvf} and {\vf} to complete execution. The x-axis lists the query graphs in decreasing order of the time taken for each to be found by {\vf}. In almost all cases, {\tempvf} outperforms {\vf}.}
	\label{fig:raw_data}
\end{figure}

There is quite a marked difference in the processing times required for {\tempvf} and {\vf}. In \figurename{\ref{fig:raw_data}} we compare the time taken for {\tempvf} and {\vf} to complete on four of the networks in our evaluation, at 30\% of the largest $d$-value in the range, for each respective network. The experiments were performed on a Linux server with a 2 GHz processor, limited to 5GB of physical memory.

The plots in \figurename{\ref{fig:raw_data}} compare the runtimes of {\tempvf} and {\vf}. In each case, the runtime of {\tempvf} is almost always faster than that of {\vf}, sometimes by orders of magnitude. In order to study the extent to which {\tempvf} outperforms {\vf}, we measure the ratio of the two sets of execution times in the following section.

\subsection{Analyzing the Speedup}

Since the difference between the time taken for {\tempvf} and {\vf} to complete can be quite large, we use the quotient of the two to compare their performance. The speedup is dependent on the $d$-value chosen. The maximum speedup happens when $d$ is at its lowest, since the pruning power is highest. As $d$ increases, the pruning is less severe, and the situation becomes increasingly like that of ordinary subgraph isomorphism. In some cases, the speedup plateaus as $d$ increases.

\begin{table}[]
	\begin{center}
		\begin{tabular}{p{2cm} rrrrrr} \hline
			Network & 10\% & 20\% & 30\% & 40\% & 50\% & $d\_max$ \\ \hline
			Prosper & 20 & 16 & 14 & 11 & 9 & 20 days \\
			Epinions & 18 & 12 & 8 & 6 & 5 & 20 days \\
			Slashdot & 6 & 5 & 5 & 5 & 4 & 30 days \\
			UC Irvine & 5 & 4 & 4 & 3 & 3 & 100 days \\
			Enron & 5 & 3 & 3 & 3 & 3 & 100 days \\
			Facebook & 1 & 1 & 1 & 1 & 1 & 300 days \\
			Hagelloch & 1 & 1 & 1 & 1 & 1 & 20 days \\
			WhoSampled & 1 & 1 & 1 & 1 & 1 & 10 years \\ \hline
		\end{tabular}
	\end{center}
	\caption{For each network and $d$-value (as a percentage of $d\_max$), the average speedup ($s$) gained from using {\tempvf} rather than {\vf}.}
	\label{tab:d_speedup}
\end{table}

\begin{figure}[]
	\centering
		
	\subfloat[Prosper]
		{\label{fig:speedup_prosper}\includegraphics[width=0.47\textwidth, height=0.25\textheight]{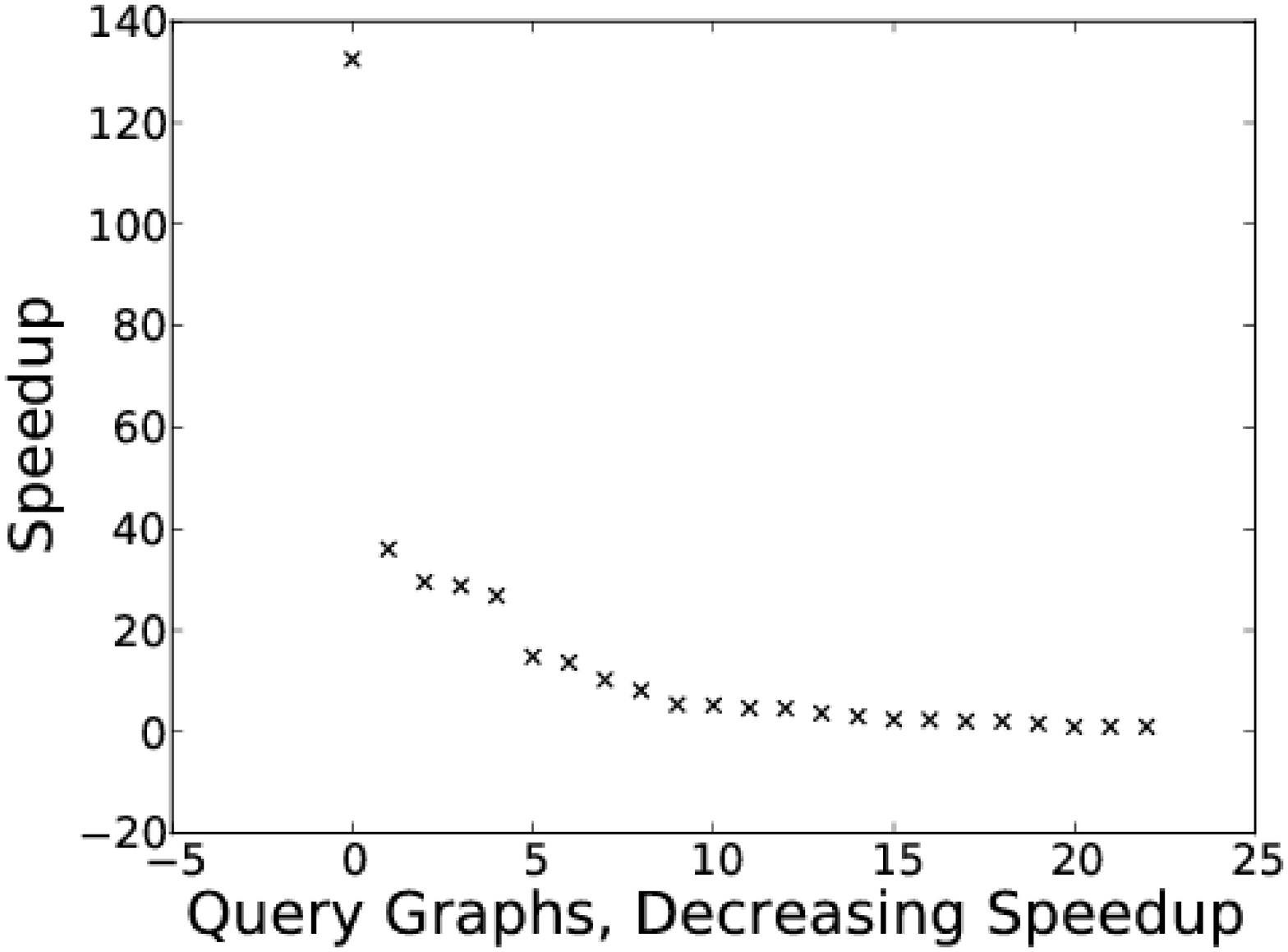}}
		\qquad
	\subfloat[Epinions]
		{\label{fig:speedup_epinions}\includegraphics[width=0.47\textwidth, height=0.25\textheight]{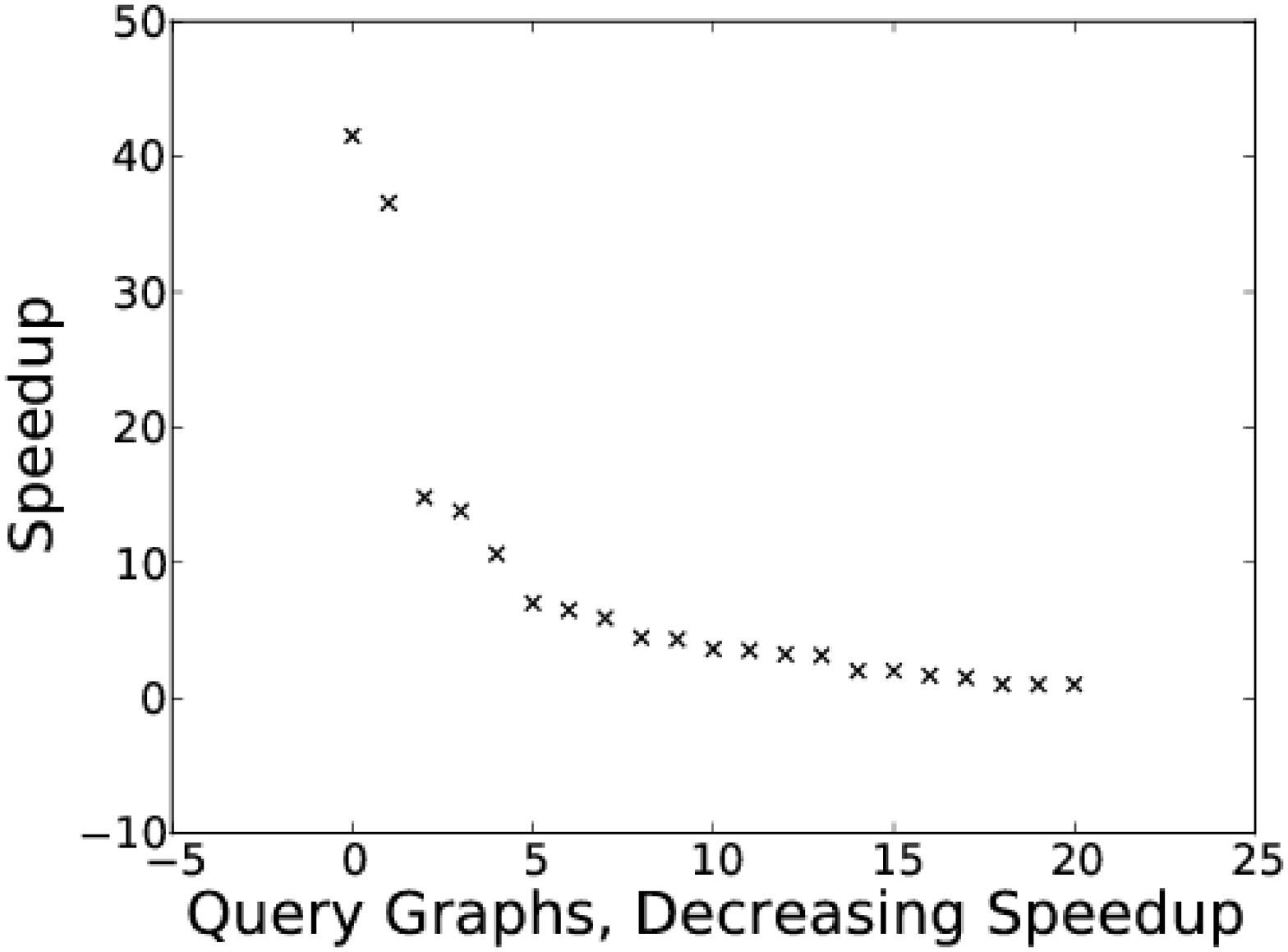}}\\
		
	\subfloat[UC Irvine]
		{\label{fig:speedup_opsahl}\includegraphics[width=0.47\textwidth, height=0.25\textheight]{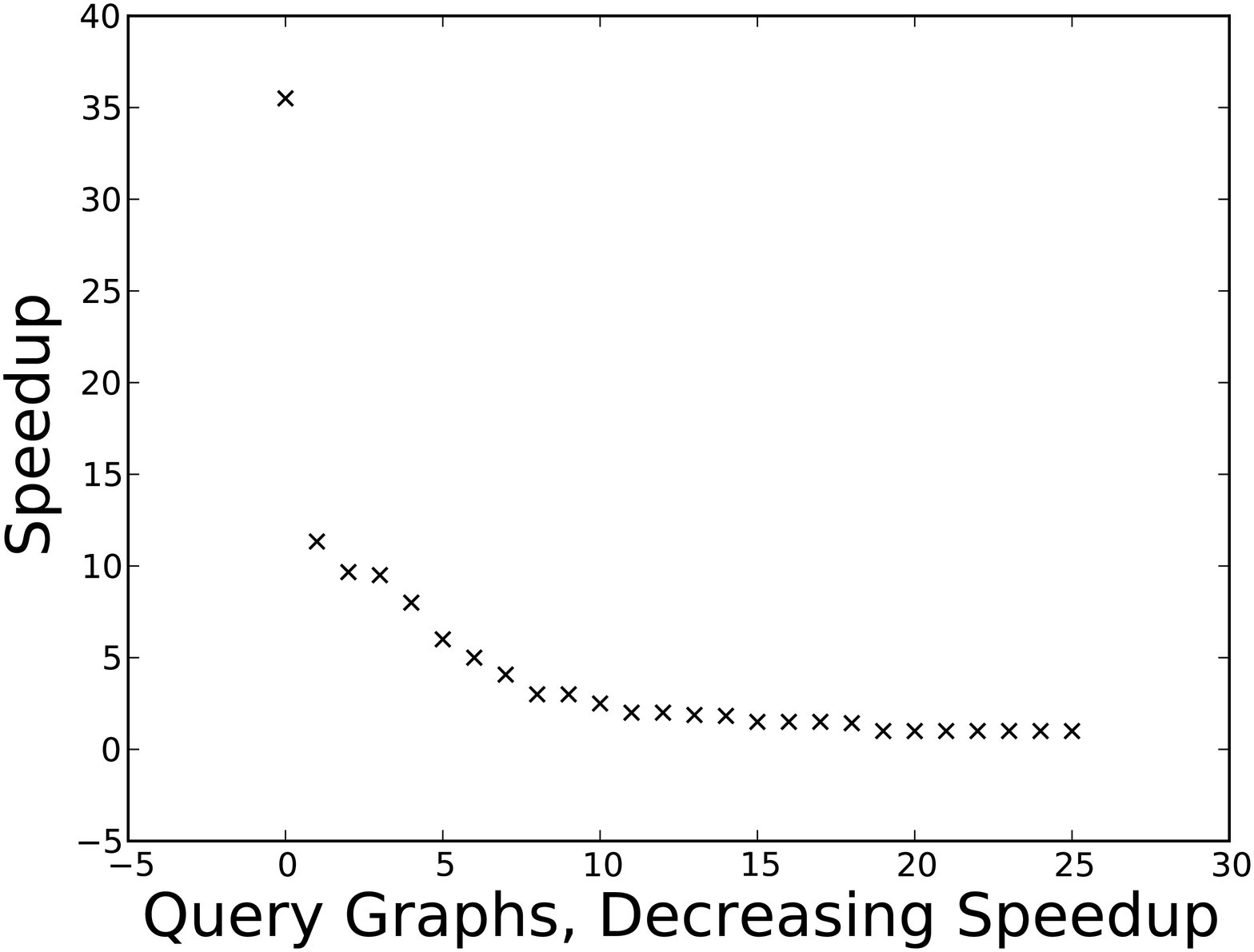}}
		\qquad
	\subfloat[Slashdot]
		{\label{fig:speedup_slashdot}\includegraphics[width=0.47\textwidth, height=0.25\textheight]{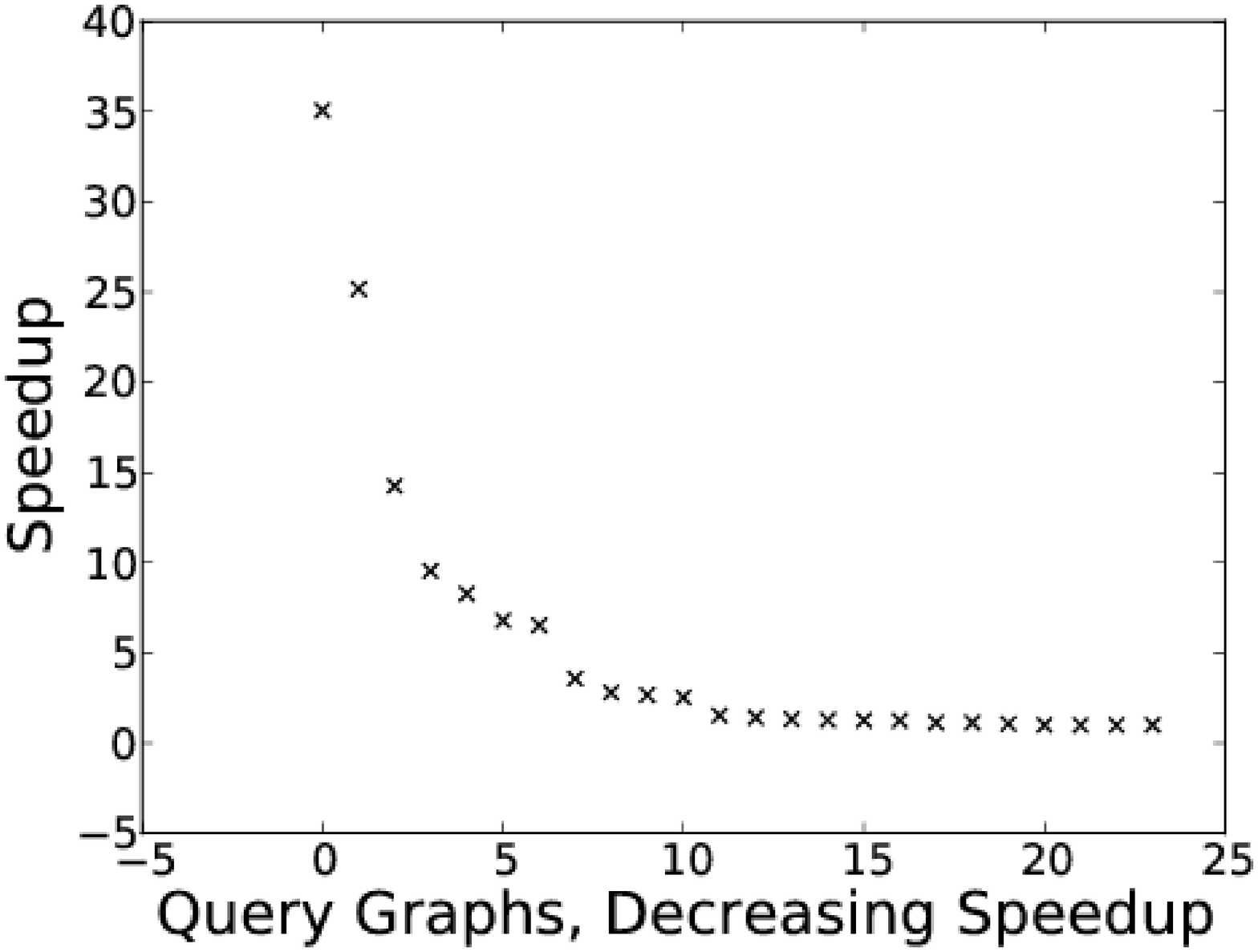}}
		
	\caption{The speedup gained by using {\tempvf} instead of {\vf} for four of the networks we studied. The comparison was done using the 30\% $d$-value appropriate for each network. Each point corresponds to a specific query graph. The x-axis ranks the query graphs in order from the query for which {\tempvf} outperformed {\vf} most significantly, to that for which the speedup was least significant. Identifying which query graphs allow {\tempvf} to outperform {\vf} reveals which scenarios {\tempvf} is most useful for.}
	\label{fig:d_speedup}
\end{figure}

The average speedup for each $d$-value is shown in Table \ref{tab:d_speedup}. As one might expect, the speedup is minimal for the naturally time-respecting networks. Since these networks are already time-respecting for very large $d$, the amount of necessary pruning is reduced. The speedup attained for each query graph at 30\% of $d\_max$ for four networks is illustrated in \figurename{\ref{fig:d_speedup}}. The speedup is shown in decreasing order from the query for which {\tempvf} most significantly outperformed {\vf}.

\begin{figure}[]
	\centering
	
	\subfloat[Prosper]
		{\label{fig:win_prosper}\includegraphics[width=0.3\textwidth, height=0.1\textheight]{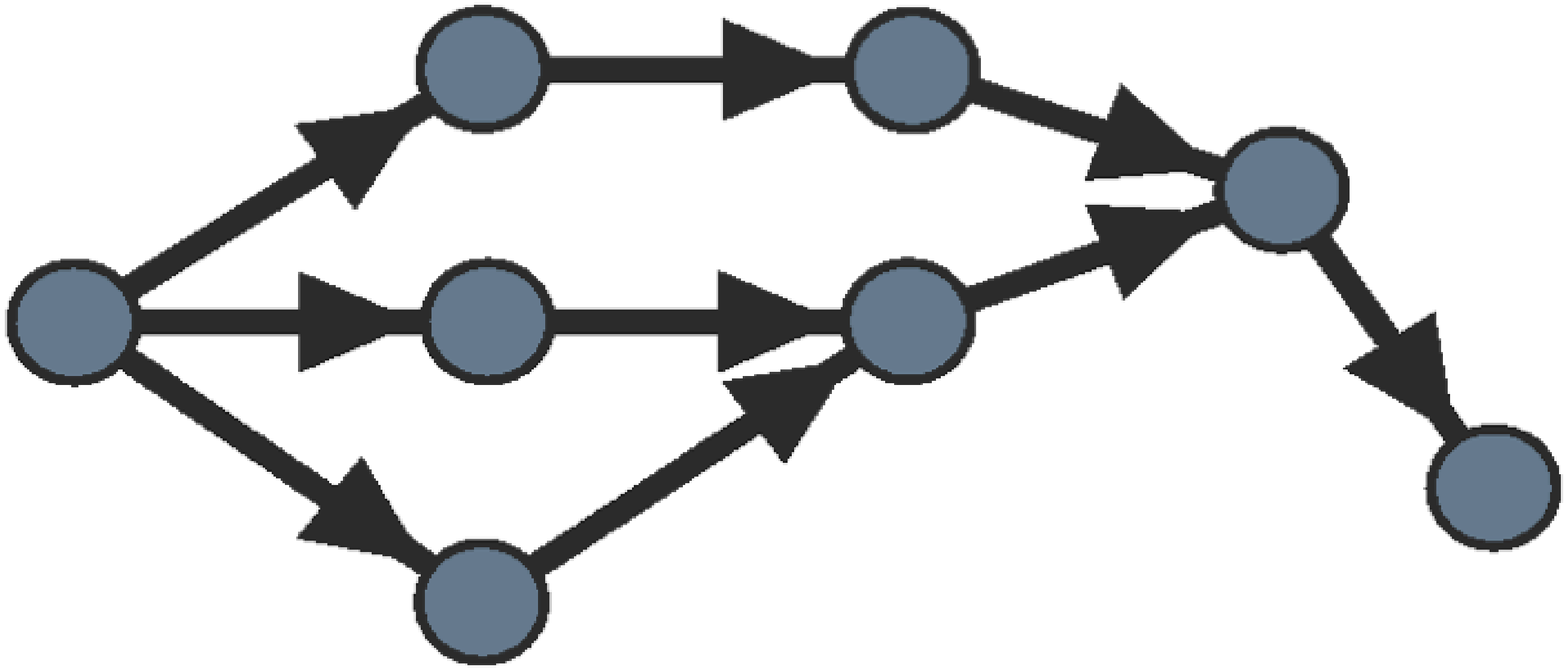}}
	\qquad
	\subfloat[Epinions]
		{\label{fig:win_epinions}\includegraphics[width=0.28\textwidth, height=0.1\textheight]{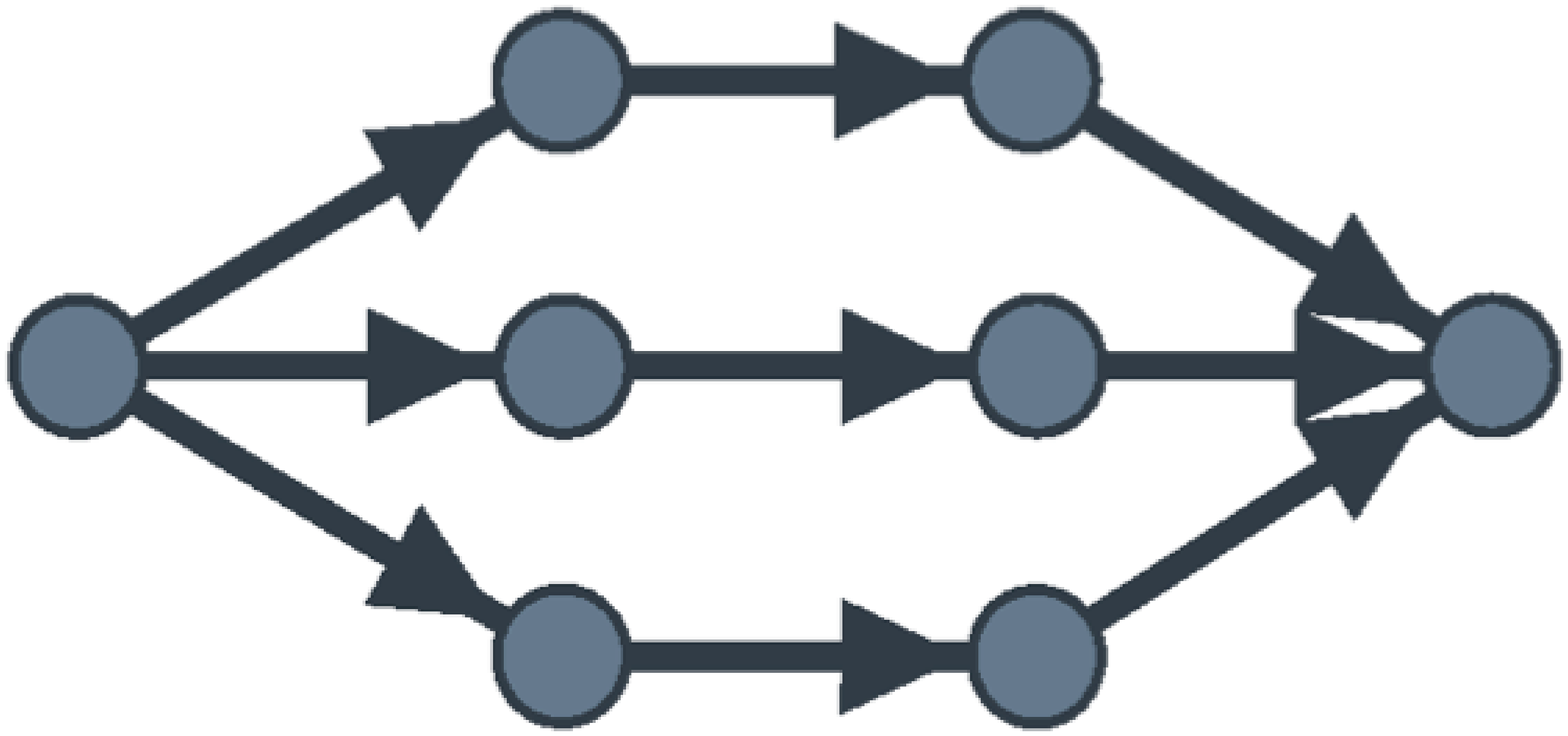}}\\
		
	\subfloat[UC Irvine]
		{\label{fig:win_opsahl}\includegraphics[width=0.26\textwidth, height=0.14\textheight]{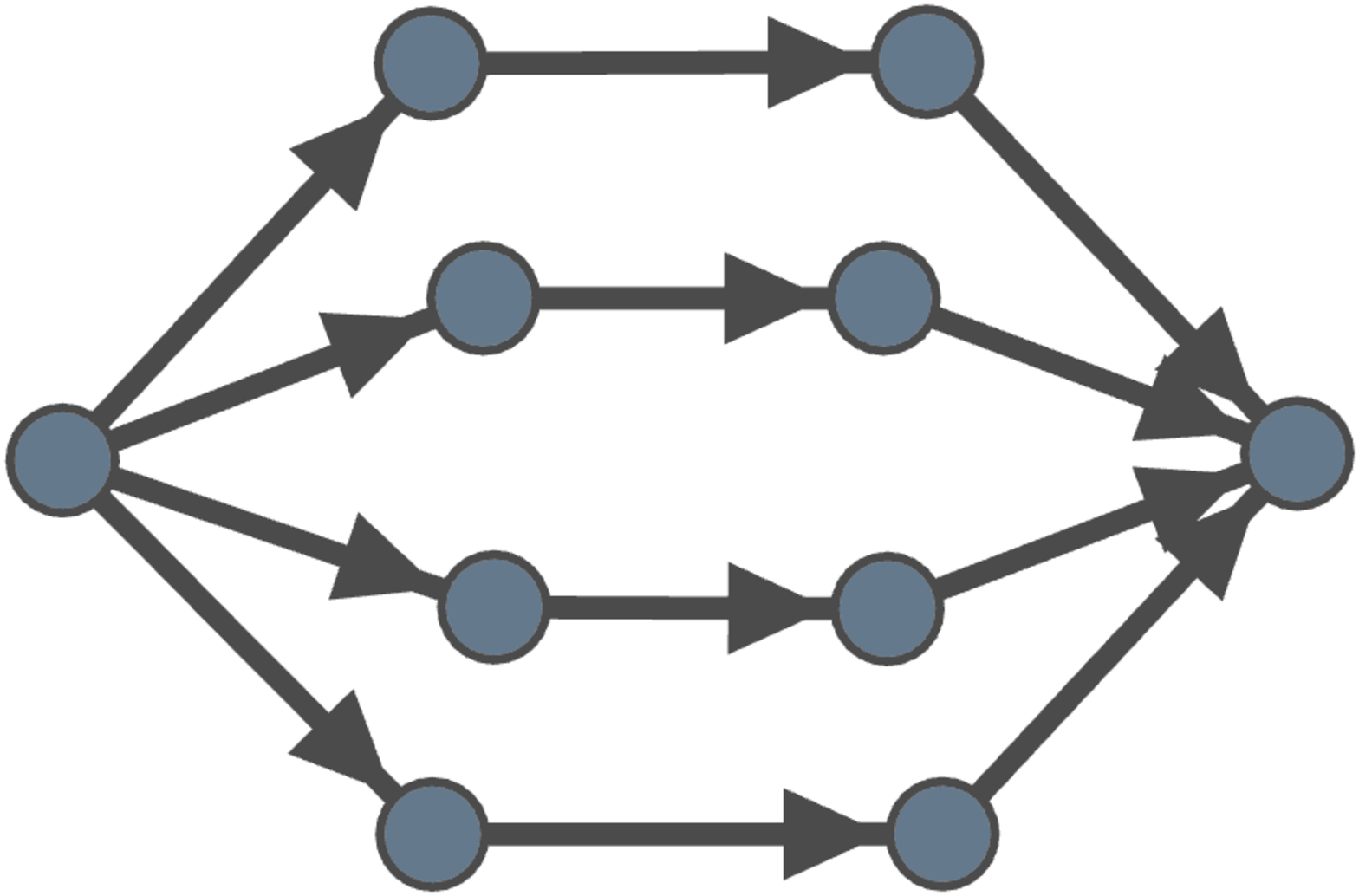}}
	\qquad
	\subfloat[Slashdot]
		{\label{fig:win_slashdot}\includegraphics[width=0.3\textwidth, height=0.06\textheight]{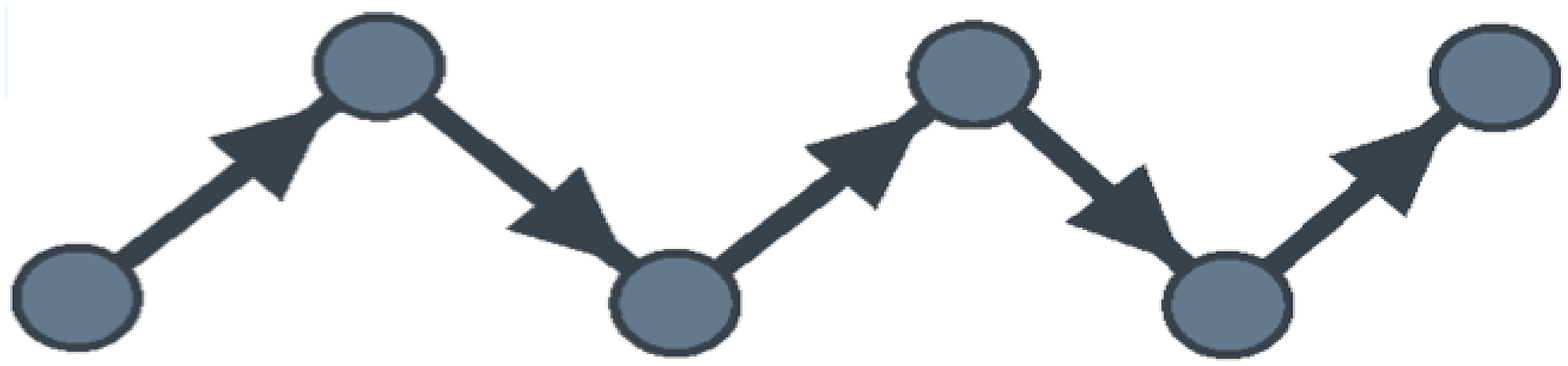}}

		\caption{The topology of the specific query graphs for which {\tempvf} outperformed {\vf} most significantly for each network in \figurename{\ref{fig:d_speedup}}. These query graphs have longer paths than the other types of graph we sought, which may be a contributing factor. Path length is related to the diameter of the query graph, discussed further in Subsection \ref{sec:factors}.}
	\label{fig:win}
\end{figure}

In all cases, the smallest performance difference occurred when the query graph was a 3-clique, while the largest difference occurred with query graphs with longer paths. The four query graphs for which {\tempvf} ran most significantly faster than {\vf} are shown in \figurename{\ref{fig:win}}. These graphs have among the largest diameters of all the query graphs we specified.

\subsection{Factors Influencing Speedup}\label{sec:factors}

Given the variance in speedup for any given network and $d$-value, it would be useful to know what factors influence this. These may be properties associated with the query graph sought or the network being analyzed. Since our results suggest that the length of paths in query graphs -- related to the diameter -- may be a factor, we study this property further.

The graph query processing literature \citep{yan-04} shows that query response time is also a major issue. This is the time taken for a search algorithm to find a query subgraph from a large graph or a database of graphs, which may use an index to aid the search. The response times of both approaches to our problem can be analyzed from a similar point of view.

The time taken for the method we aim to improve on may be specified as follows.

\begin{displaymath}
	(\vf)_{cost} = T_{search} + |C_{q}| \times T_{temporal\:test}
\end{displaymath}

First, the search process must complete without pruning, captured by $T_{search}$. This returns a set of candidates, $|C_{q}|$ which topologically correspond to the query graph. Each must be tested, as in $T_{temporal\:test}$, to ensure that it meets the time-respecting requirements, and only those that do are retained. Thus, the number of tests depends on the number of candidates. The time taken for our more efficient method may be described as follows.

\begin{displaymath}
	(\tempvf)_{cost} = T_{temporal\:search}
\end{displaymath}

Using this method, no candidates are generated. Only $T_{temporal\:search}$ is computed. As an embedding is expanded, the test for the time-respecting property is performed. If the property is violated, the embedding is immediately discarded. This avoids using up memory with erroneous candidates, and correspondingly removes the need to test potentially large sets of candidates.

The instances of $|C_{q}|$ are displayed in Table \ref{tab:spur_cand}. As expected, $|C_{q}|$ is large when $d$ is small, and reduces as $d$ increases. This is explained by the pruning power of $d$ -- the more pruning that occurs, the more candidates are discarded. As suggested previously, $|C_{q}|$ is small for the naturally time-respecting networks. The communication, social and transaction networks offer the greatest potential for speedup, since they provide the largest $|C_{q}|$ values. This is reflected in the plots of \figurename{\ref{fig:d_speedup}}.

\begin{figure}[]
	\centering
	
	\subfloat[UC Irvine spurious candidates.]
		{\label{fig:opsahl_candidates}\includegraphics[width=0.47\textwidth, height=0.24\textheight]{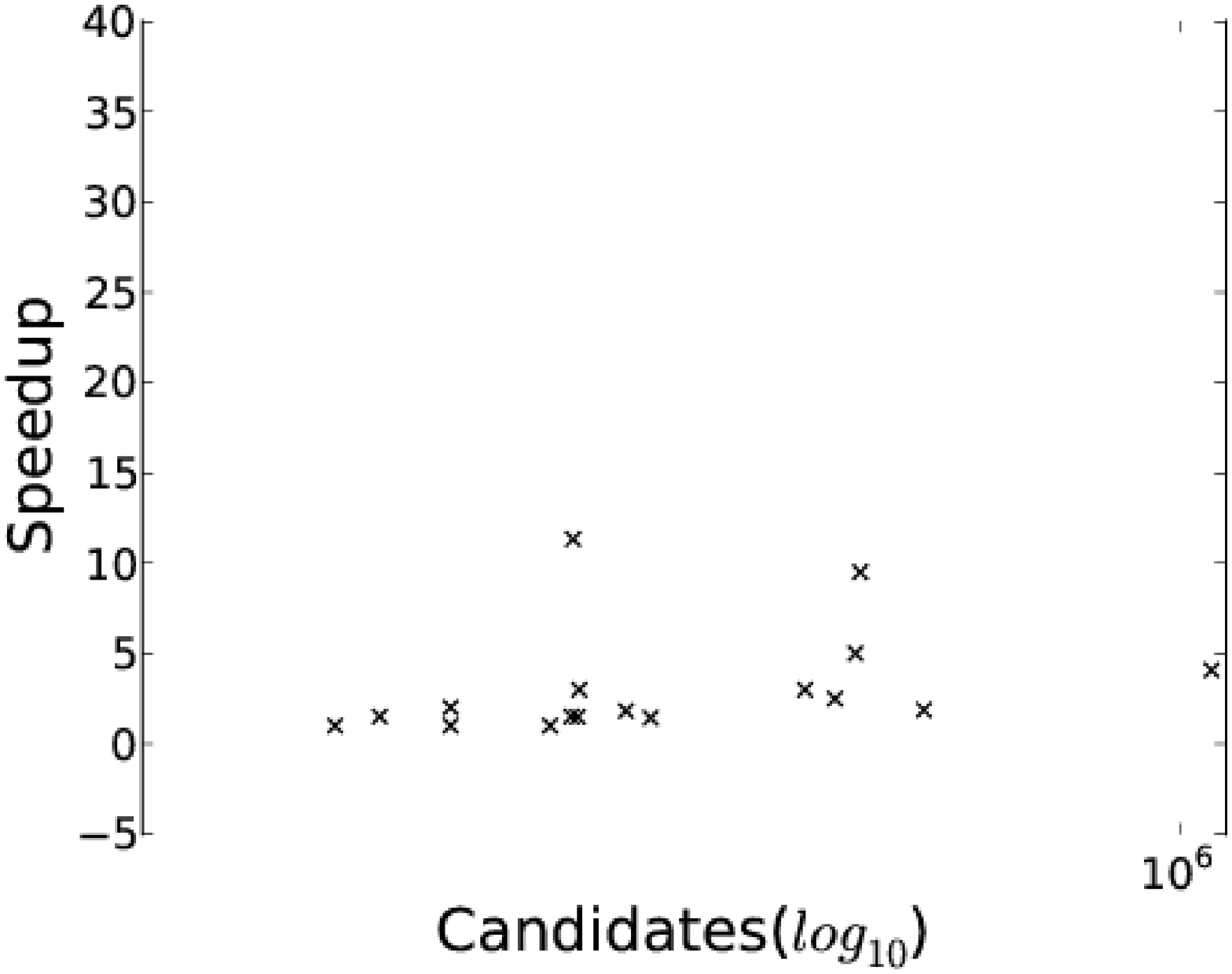}}
		\qquad
	\subfloat[Slashdot spurious candidates.]
		{\label{fig:slashdot_candidates}\includegraphics[width=0.47\textwidth, height=0.24\textheight]{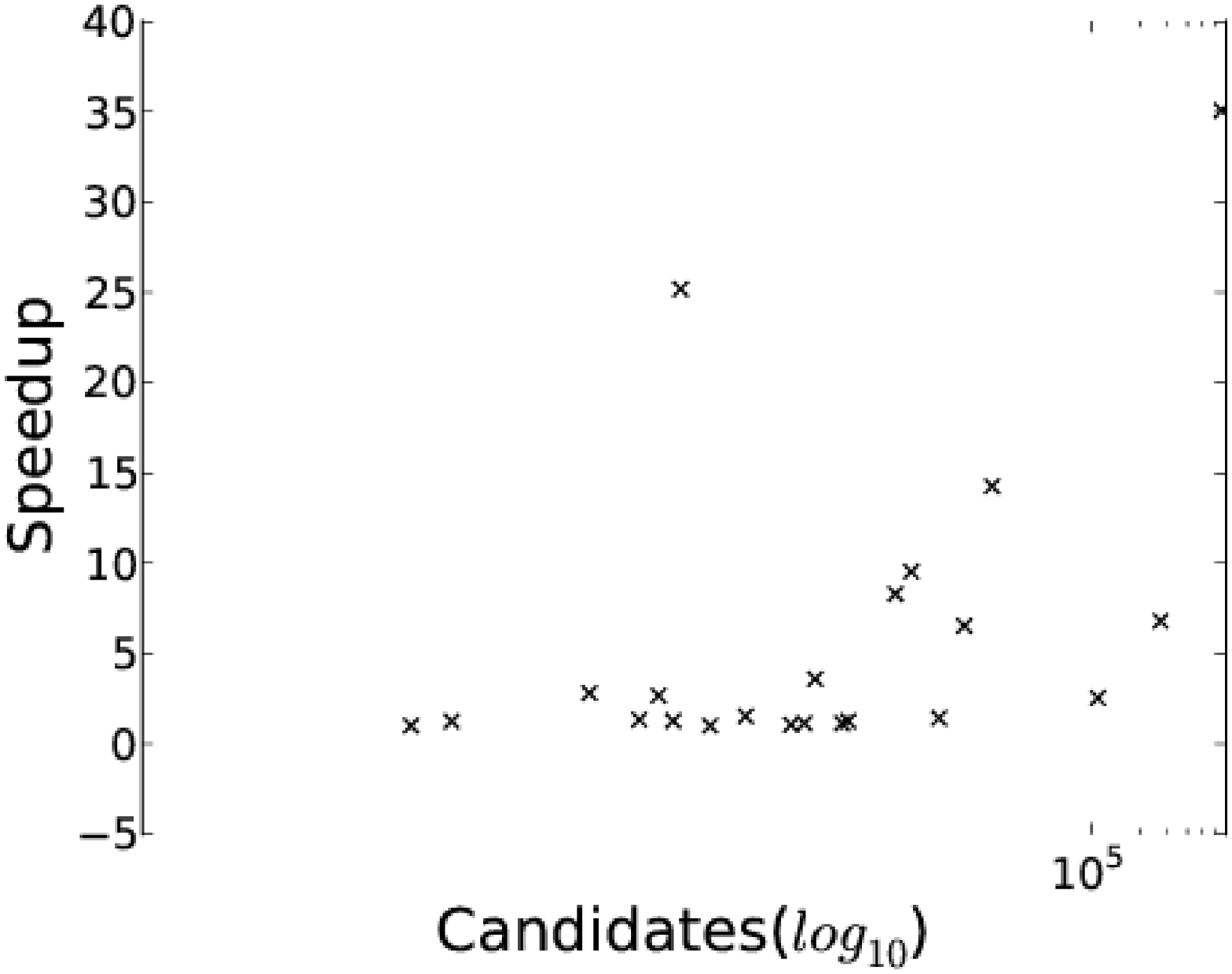}}\\
		
	\subfloat[UC Irvine diameter.]
		{\label{fig:opsahl_diameter}\includegraphics[width=0.47\textwidth, height=0.24\textheight]{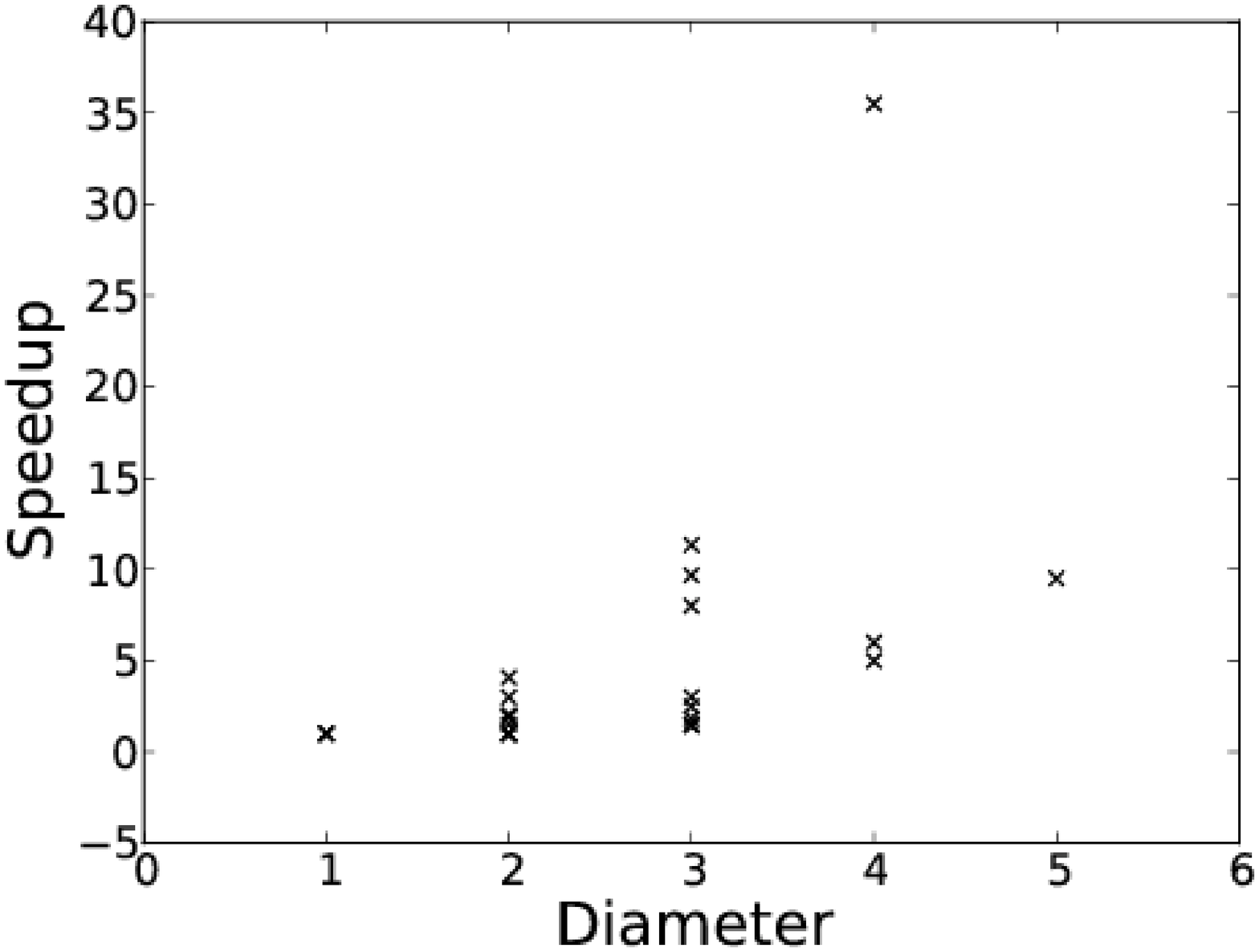}}
		\qquad
	\subfloat[Slashdot diameter.]
		{\label{fig:slashdot_diameter}\includegraphics[width=0.47\textwidth, height=0.24\textheight]{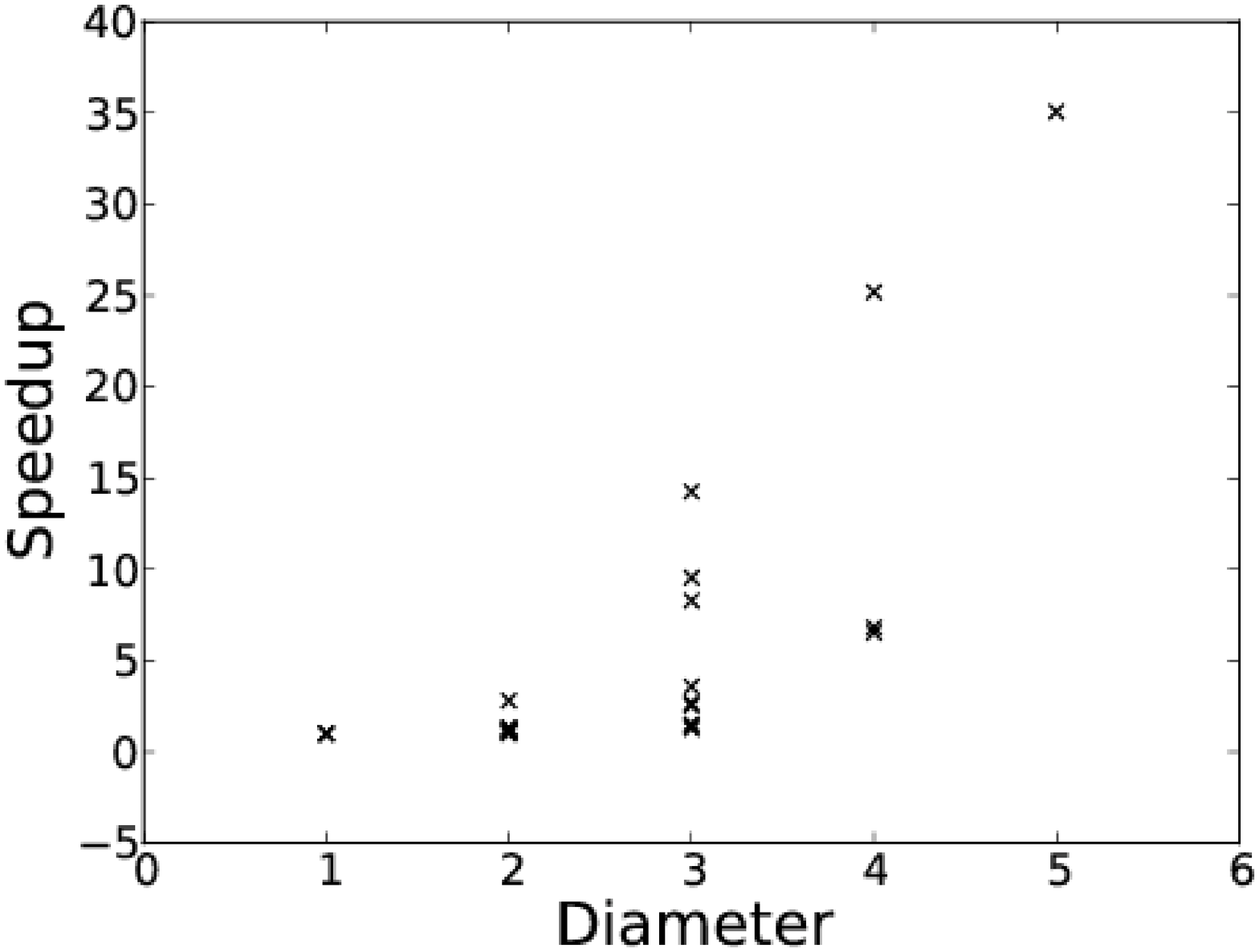}}\\
		
	\subfloat[UC Irvine size.]
		{\label{fig:opsahl_size}\includegraphics[width=0.47\textwidth, height=0.24\textheight]{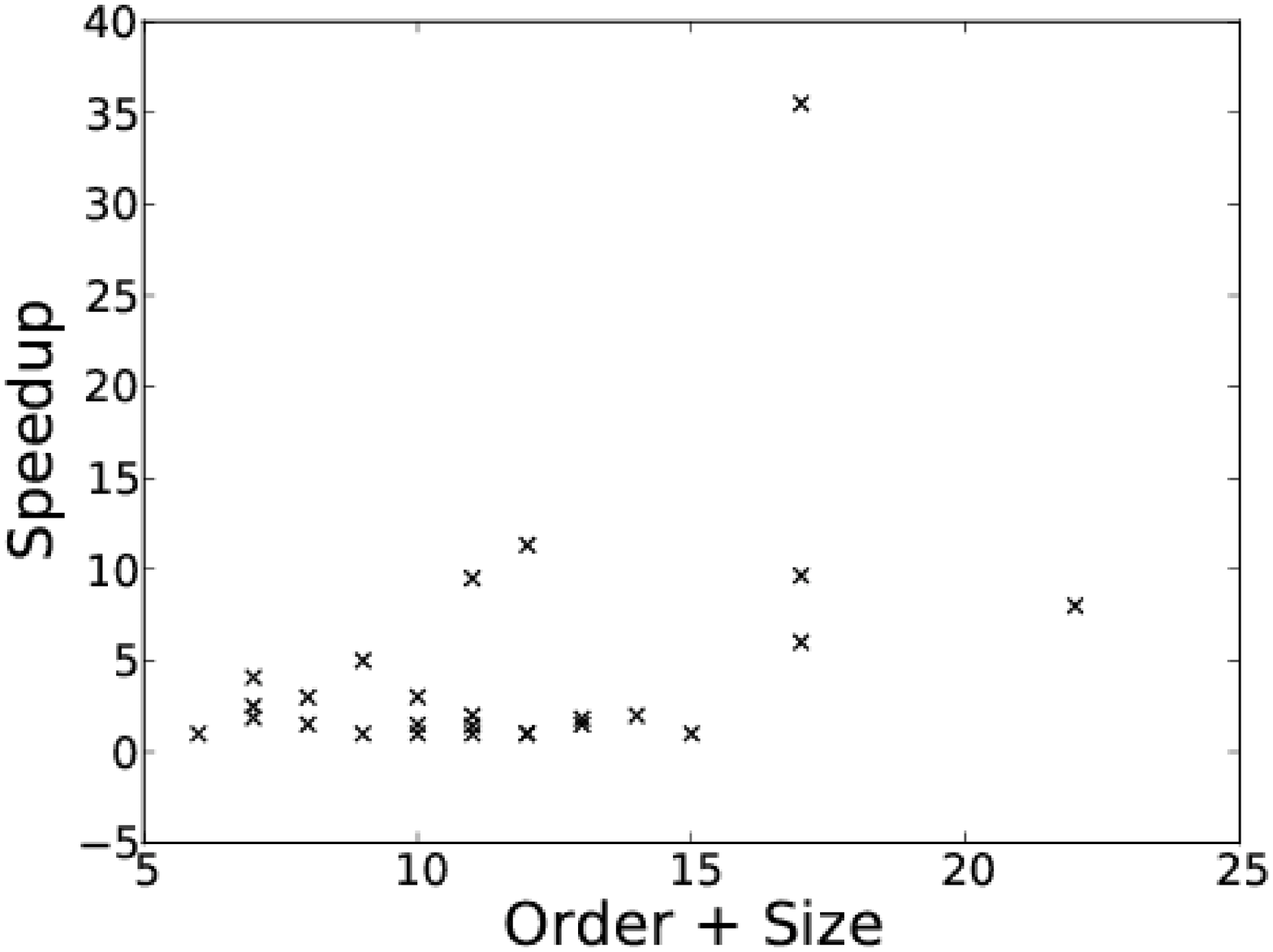}}
		\qquad
	\subfloat[Slashdot size.]
		{\label{fig:slashdot_size}\includegraphics[width=0.47\textwidth, height=0.24\textheight]{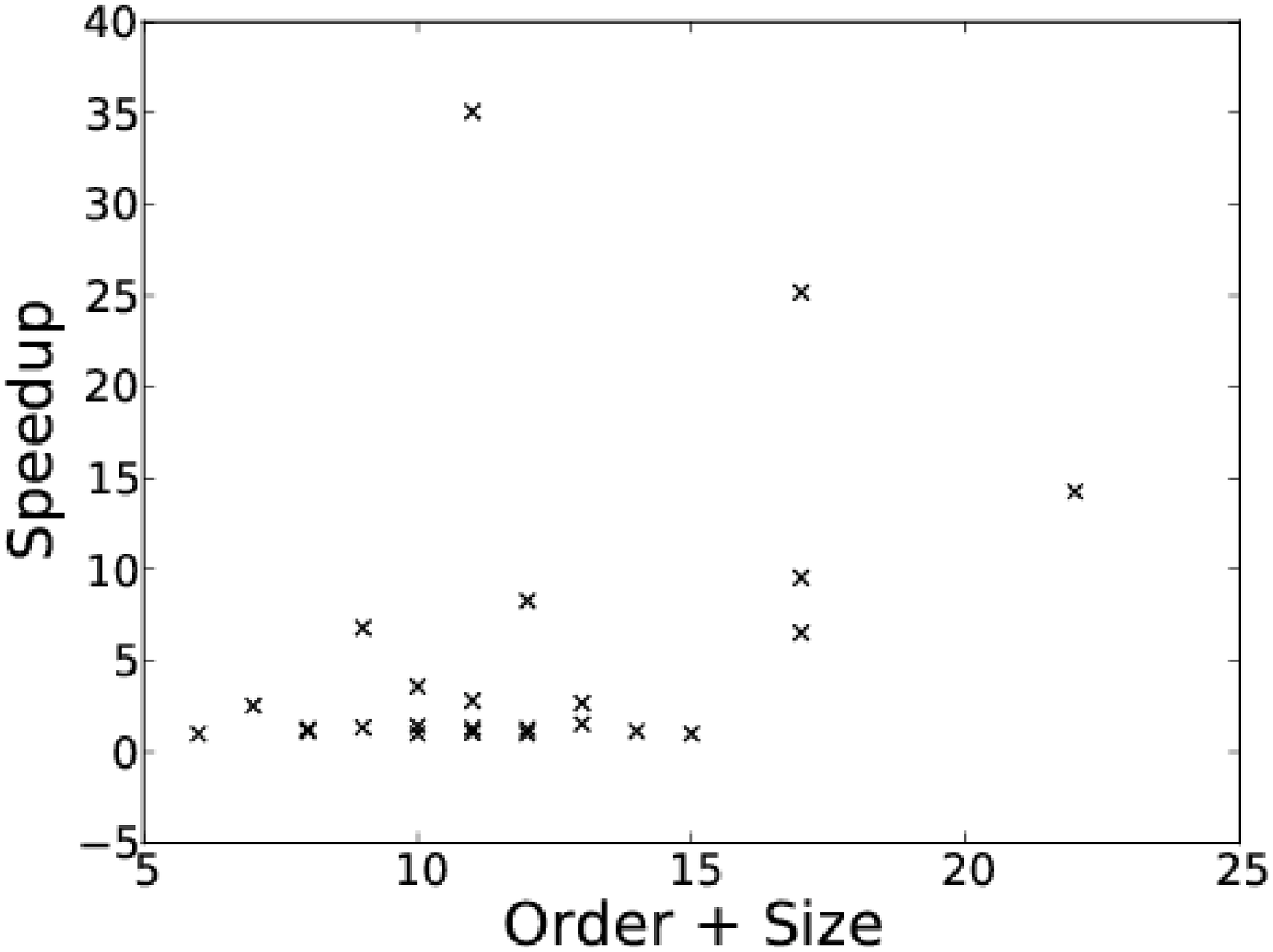}}\\
		
	\caption{Predictors of speedup. In all cases, the y-axis represents the ratio of {\vf} to {\tempvf}. (\ref{fig:opsahl_candidates}, \ref{fig:slashdot_candidates}) The {\vf} approach generates many spurious candidate embeddings, which turn out not to be time-respecting. Thus, as the number of spurious candidates increases along the x-axis (on a log scale), the extent to which {\tempvf} outperforms {\vf} also increases. (\ref{fig:opsahl_diameter}, \ref{fig:slashdot_diameter}) As the diameter of the query graph increases along the x-axis, {\tempvf} performs better than {\vf}. (\ref{fig:opsahl_size}, \ref{fig:slashdot_size}) As the size of the query graphs increases along the x-axis (computed as the sum of the nodes and interactions), so too does the speedup gained from using {\tempvf} instead of {\vf}.}
	\label{fig:predictors}
\end{figure}

Although $|C_{q}|$ is a strong predictor of speedup, other factors come into play. The features of the query subgraphs that we found to have the most influence over the speedup are the diameter -- evaluated by viewing the subgraphs as undirected -- and the total size -- which for the evaluation we define as the sum of the vertices (order) and interactions (size). As shown in \figurename{\ref{fig:predictors}}, these features to some extent help to predict the likelihood of speedup gained by using the {\tempvf} method rather than {\vf}. This result is corroborated by the query graphs illustrated in \figurename{\ref{fig:win}}. These all have larger diameters and size than most other query graphs sought. 

\begin{table}[]

\begin{center}
\begin{tabular}{lrrrrrr}\hline
	Network & 10\% & 20\% & 30\% & 40\% & 50\% & $d\_max$ \\ \hline \hline
	Enron & \numprint{75996} & \numprint{72846} & \numprint{72346} & \numprint{72329} & \numprint{65947} & 100 \\
	Facebook & \numprint{335453} & \numprint{286061} & \numprint{269084} & \numprint{239588} & \numprint{194840} & 300 \\
	UC Irvine & \numprint{1173003} & \numprint{1009080} & \numprint{893075} & \numprint{764935} & \numprint{689266} & 100 \\
	Slashdot & \numprint{1114942} & \numprint{1113895} & \numprint{1113095} & \numprint{1112373} & \numprint{1111356} & 30 \\ \hline
	
	Epinions & \numprint{1477219} & \numprint{1467752} & \numprint{1456123} & \numprint{1444498} & \numprint{1428721} & 20 \\ \hline

	Hagelloch & \numprint{24469} & \numprint{24207} & \numprint{23496} & \numprint{22236} & \numprint{21739} & 20 \\
	WhoSampled & \numprint{9312} & \numprint{8404} & \numprint{5514} & \numprint{4566} & \numprint{3584} & 10 \\ \hline

	Prosper & \numprint{1885825} & \numprint{1869491} & \numprint{1842836} & \numprint{1798623} & \numprint{1753719} & 20 \\ \hline
\end{tabular}
\end{center}

\caption{For each network and $d$-value (as a percentage of $d\_max$), the number of spurious candidates generated. This has a large bearing on query response time. As the count of spurious candidates increases, so too does the extent to which {\tempvf} outperforms {\vf}. This happens because {\tempvf} does not generate spurious candidates.}
\label{tab:spur_cand}
\end{table}

\section{Conclusions and Future Work}\label{sec:conc}

The solution of the subgraph isomorphism problem has important applications for many network types. Since the problem is NP-complete, the solution becomes intractable as the network size grows. When the network in question contains temporal information, however, the problem can be constrained. Instead of searching for embeddings of a query graph with only the topology specified, we also insist that the embeddings are time-respecting. In this work, we evaluated three related methods to solve this problem. The approach that first extracted all time-respecting subgraphs and then performed subgraph isomorphism testing proved too computationally expensive and inefficient to be reasonable. Of the two fastest methods, {\tempvf} performs faster than {\vf} on all types of network studied, and sometimes by a significant factor.

The benefits of the {\tempvf} method were more pronounced when the query graphs were composed of longer paths. Thus, in scenarios where query graphs with long paths which are time-respecting are sought, {\tempvf} will return results quickest. On the other hand, it will take about as long to find graphs with shorter paths, such as cliques, if either {\tempvf} or {\vf} are used.

Our approach was most effective in the context of the social, communication and financial transaction networks we studied. The number of spurious candidates generated by {\vf} may explain this. Since there are a large number of topological matches for the query graphs sought within these networks, many potential matches are found which turn out not to be time-respecting. These matches are not generated by {\tempvf}, which checks that the match is time-respecting earlier in the search. In the case of the naturally time-respecting networks and communication networks, there were fewer extra candidates generated by {\vf}. A more detailed theoretical complexity analysis could yield further insights into the relationship between the performance of our temporal approach and the properties of the data graph and query graph. This is an important direction for future work.

This work has focused on the exact matching problem, insisting that the topology of an embedding exactly matches that of the query graph. Since the embedding is induced, prospective matches will not be returned if there are extra interactions in the embedding, for example. Also, embeddings that are similar to the query graph are not returned, although they might be interesting for the user. In future, we aim to extend this work to include inexact matching. This will also allow us to deal more effectively with networks that might have missing data.

Another consideration relates to bursty interactions between groups of individuals over time. If a group interacts intensively over the space of an hour, and this occurs once every week, it might be interesting to first time-slice the data following a dynamic network approach, and then perform {\tempvf} on the time-slices.

In the current work, we have restricted the size of the data graphs to \numprint{20000} interactions. This has demonstrated that the temporal approach is more feasible than the static approach. It would be interesting to see how the temporal model scales to much larger networks -- which are increasingly available for study -- and larger query graphs. Memory requirements could have a large bearing on the scalability. A study of memory usage during the matching processes for the graphs used in this work would provide an indicator of the viability of the algorithm. An alternative implementation, for example an extension to the efficient AllDifferent-based filtering algorithm \citep{solnon-10}, may make the matching process faster still. If the time-respecting requirement is viewed as a constraint, some interesting comparisons with constraint programming approaches could be made, for example Choco \citep{choco} or Gecode \citep{gecode}.

An extension of the algorithm to deal with graph data arriving in a stream would also be a worthwhile direction. With nodes and interactions arriving over time, an incremental approach would be required, since recomputing matches repeatedly would not be feasible.

\section*{Acknowledgements}

This work was supported by Science Foundation Ireland [08/SRC/I1407, SFI/12/RC/2289]. 

\section*{References}

\bibliography{mybibfile}

\end{document}